\documentclass{article}[12pt]
\usepackage[latin1]{inputenc}
\usepackage{graphics,graphicx,longtable,verbatim}
\usepackage{setspace}
\usepackage{amsmath}
\usepackage{amssymb}
\usepackage{bbm}
\usepackage[english]{babel}
\usepackage{hyperref,float}
\usepackage[authoryear]{natbib}
\usepackage[top=1.5in, bottom=1.5in, left=1in, right=1in]{geometry}
\usepackage[]{algorithm2e}
\usepackage{authblk}

\title{Inferring learners' affinities from course interaction data}

\author{Maria Osipenko} \affil{Berlin School of Economics and Law, Berlin, Germany\\
osipenko@hwr-berlin.de
 }
\date{}

\providecommand{\keywords}[1]
{
  \small	
  \textbf{\textit{Keywords---}} #1
}

\begin{document}
\maketitle
\begin{abstract}
A data-driven model where individual learning behavior is a linear combination of certain stylized learning patterns scaled by learners' affinities is proposed. 
 The absorption of stylized behavior through the affinities constitutes "building blocks"  in the model.  Non-negative matrix factorization is employed to extract common learning patterns and their affinities from data ensuring meaningful non-negativity of the result.
The empirical learning patterns resulting from the actual course interaction data of 111 students are connected to a learning style system. Bootstrap-based inference allows to check the significance of the pattern coefficients. Dividing the learners in two groups "failed" and "passed" and considering their mean affinities leads to a bootstrap-based test on whether the course structure is well-balanced regarding the learning preferences.
\end{abstract}

%%%%%%%%%%%%%%%%%%%%%%%%%%%%%%%%%%%%%%%%
%\vspace{10cm}
\section{Introduction}
Structured assessment of learning preferences in an online learning framework builds a basis for course improvement and offers possibilities for creating a productive learning environment with a potentially pronounced effect on the learning quality as e.g. \cite{ford}. Informing learners about their learning styles based on their course interaction data may enhance their awareness of the materials that suit best for their learning progression (\cite{graf2}). Though the individual learning strategies vary, they are assumed to share some common components, termed learning styles (\cite{feldman}). These styles are expected to be reflected in the actual course interactions through stylized {\it patterns} in the data, characteristic for each learning style. The resulting stylized behavior is absorbed through individual preferences. When linked to behavioral patterns from the observable learners' data, those preferences are parameterized by what I term here as {\it affinities} towards each pattern. These individual affinities measure how pronounced the stylized pattern is reflected in the observed learner's behavior.

%course environment
In this paper I propose a coherent data-based model, where individual learning behavior is represented as a linear combination of certain stylized learning patterns personalized through individual affinities. The absorption of stylized behavior through these affinities constitutes in the model "building blocks" for individual learning behavior and facilitates the understanding of how the learning styles could conceptually "flow in" the overall personalized learning process. The proposed model is consistent with the widely recognized learning style model of \cite{felder}, where according to \cite{felder2} learners' behavior may possibly reflect all learning styles with different strengths due to its time- and context-varying nature.
It is useful to connect the estimated stylized patterns to a coherent learning style model to derive recommendations for course improvement based on the model estimated from data.

%theories
There exist  a variety of learning style models that potentially build such a basis, see \cite{feldman} for a survey thereof. In this analysis, I use the most referenced (according to the authors in \cite{feldman}) learning style classification of \cite{felder}. The later incorporates four levels: processing (active/ reflective learners), perception (sensing/ intuitive), input (visual/ verbal), and understanding (sequential/ global). One more dimension regarding the learning attitude can be found in \cite{biggs} separating i.a. surface and achieving learning styles.
 
The following properties characterize the learning styles mentioned above. Whereas active learners learn by trying out, applying concepts, and tend to collaborate with others, reflective learners prefer to work along reflecting the material. Sensing learner learns best by examples and like standard tasks, intuitive style determines preference for abstraction and generalized principles. Visual versus verbal styles regard the form of the learning material preferred. Step-by-step learning identifies sequential learner and the need of big picture and the vice versa characterizes the global type.  Finally, low engaged learners or surface learners, put only enough afford to complete or pass the course, whereas the achievers desire highest grades (\cite{feldman}).

The models of learning styles provide questionnaires to determine the learning preferences through the explicit feedback of the learners. 
 This approach was termed collaborative by \cite{brusilovski} and is criticized as inconsistent, e.g. in \cite{Kirschner}.
The automatic approach as \cite{brusilovski} aims at identifying the learning styles based on the actual behavior.  \cite{sheeba} distinguishes, thereby, a data-driven and a literature-based approaches. In the data-driven setting learning styles are extracted from the data using an appropriate algorithm, which combines the input data to output learning styles by optimizing its loss function. Literature-based approaches give simple rules on calculating learning styles from particular behavioral features (see e.g. \cite{lwande}, \cite{graf}). In this paper I follow the data-driven approach. The resulting learning patterns are then related to the learning styles mentioned above based on their behavioral characteristics.

%literature review on extraction of learning styles
 Many authors, who attempt to extract behavioral patterns corresponding to the learning styles in the data-driven setting, use {\it supervised} techniques such as classification to obtain a model matching the actual behavior to the prespecified learning style labels, obtained through questionnaires, interviews or expert opinions. For instance, \cite{sheeba} determine the learning styles based on the guidance of \cite{felder} for the observed learners and use decision trees to combine the observed features in order to forecast the learning preferences. \cite{GARCIA2007} employ  questionnaires  to match the learning style and use Bayesian networks to fit a predictive model and to obtain the probability of each learning pattern. \cite{troussas} train an ensemble of three classifiers  and \cite{zhang} employs deep neural networks for learning style detection based on predetermined labels. I refer to \cite{wibirama} for further proposed approaches to supervised learning style prediction.

A drawback of supervised methodologies is the requirement to provide some learning style reference as ground truth, normally obtained by filling out questionnaires, interviews, self-reporting, etc. These labels are not always available or reasonable and may be error prone (\cite{hamd}, \cite{Kirschner}).

Another option is to use an {\it unsupervised} technique (also referred to as implicit modeling in \cite{hamd}) to extract the common behavioral patterns directly from the observed behavioral data based on minimizing some self-sufficient loss function (e.g. squared error of the approximation).

For instance, \cite{SHRESTHA} use k-means clustering as an intermediate step to map the data to some known learning styles.
However, k-means assign membership to only one cluster. Consequently, each learner can only belong to a single group. That is, no one-to-one relation of their clusters to the original learning styles can be established. The resulting clusters rather comprise frequent combinations of learning patterns present in a considered group of students. Transferring k-means results to another group of students may be problematic. 
\cite{azzi}overcome this problem by using fuzzy c-mean clustering for unsupervised detection of learning styles. The authors then inspect the resulting clusters and label them according to the model of \cite{felder}. Their cluster centers can be interpreted as the behavioral prototypes of the learning styles and the degrees of cluster memberships correspond to the strengths of the individual preferences.  Similar approaches is taken in \cite{faiss} and \cite{hamd}. However, the formation of individual learning behavior (as a combination of the cluster memberships) lacks comprehension in such a model. \cite{scaccia} use dimension reduction via principal component analysis (PCA) to model common students' learning patterns and monitor temporal changes therein. 
The actual behavior is approximated as a sum of principal component loadings (learning patterns) scaled by the respective scores. As a result, PCA offers a generative model for the observed learners data. However, because  both principal component scores and loadings are allowed to be positive and negative, the resulting components are sometimes added and sometimes subtracted. This fact complicates the interpretation of the result in terms of learning patterns as "building blocks" for the observed learners and possibly negative scores as individual affinities.

In this paper, I use another technique -  non-negative matrix factorization - to extract the common learning patterns and the corresponding affinities in order to ensure the non-negativity of the both. Compared to the PCA method of \cite{scaccia}, it offers the advantage that the extracted learning patterns and the preferences are non-negative and can be easily interpreted as suitably scaled additive "building blocks" determining the individual learning behavior. 

I apply the model to extract the common learning patterns and the corresponding affinities from learning interaction data of 111 students of an online statistics course. The resulting empirical learning patterns are connected to the learning styles mentioned above. It is possible to construct a statistical test using mean per-pattern affinities of two groups "failed" and "passed" to get the insights on whether the course structure is well-balanced regarding learning preferences, which is also demonstrated.

The paper is organized as follows. In the next section, I describe the statistical model for extracting persistent patterns of learning behavior and respective affinities from data and introduce non-negative matrix factorization as the underlying approach for finding the unknown parts of the model. The section, where I  briefly address the course framework, the available data, and the preprocessing steps, follows.  I also present the extracted learning patterns and the corresponding affinities and show how to use them to check whether the course structure is well-balanced to meet the needs of different learners in this section. Finally, I conclude  on the findings.

\section{Model specification and estimation}

To begin with, assume that each learning style of the adopted learning style system is reflected in the respective unambiguous {\it stylized} learning behavior, which is typical for it. Learners' course interactions arise then as a result of combining the stylized behavior patterns of their preferred learning styles. These stylized patterns are thereby common to all learners. But since learners' preferences differ, also the importance of the respective stylized patterns for the formation of their individual learning behavior vary.  I refer to the stylized behavior in further as to {\it (common) learning patterns} and to the linked individual importance as to {\it affinities} thereto.

\begin{figure}[H]
	\centering
	\includegraphics[scale=0.8]{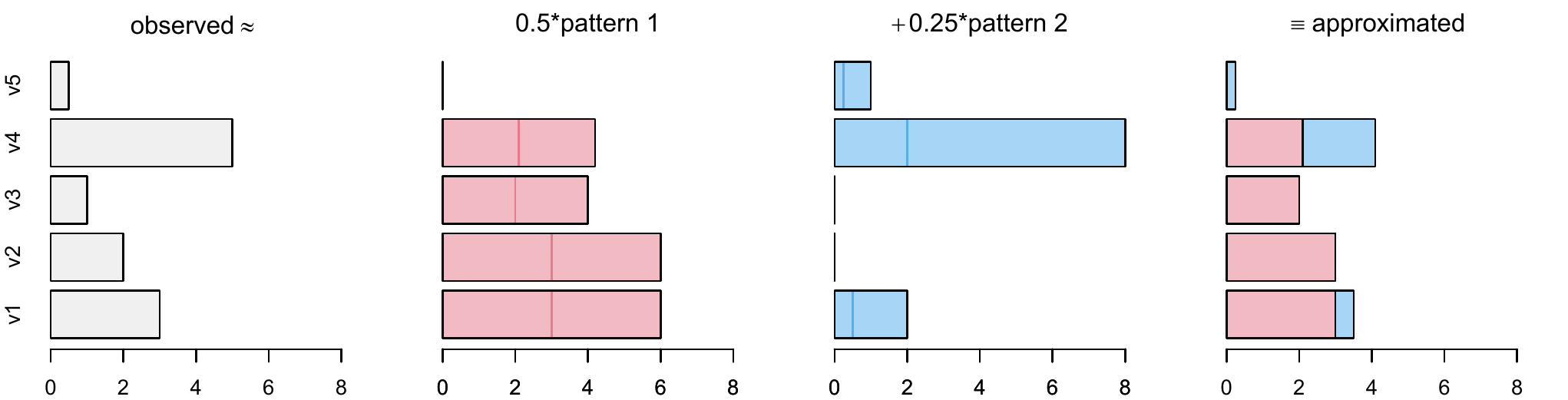}
	\caption{\scriptsize{Illustration of approximating observed behavior described by five variables $v_1,\ldots, v_5$ (left) as a linear combination of the two patterns, pattern 1 (middle left) and pattern 2 (middle right), with the respective affinities of 0.5 and 0.25. The resulting approximation is shown on the right panel.
			}}\label{pic}
\end{figure}

Moreover,  assume that the observed individual learning behavior can be approximately represented by a{ \it linear combination} of the common learning patterns according to their individual affinities.
Those patterns build, in a sense, stretchable and shrinkable blocks, which can be put on top of each other.  Each individual learning pattern in this model is composed out of these "building blocks" stretched or shrunk by individual affinity to the respective pattern as illustrated in Figure \ref{pic}.

Let $X$ further denote a $p\times n$-matrix where the columns contain $p$ observed course interaction measurements for each of $n$ learners. The aim is to approximate $X$  by a product of a non-negative $p\times K$-matrix $P$, containing the unobserved common learning {\bf p}atterns, and a transpose of a non-negative $n\times K$-matrix $A$, likewise unobserved, holding the individual {\bf a}ffinities in its columns. 
Formally, $X$ can be written as:
\begin{align}
X\approx P A^\top,\label{xnmf}
\end{align}
where all elements of $P$ and $A$ are non-negative. The non-negativity constraint ensures that both the learning patterns and the individual affinities thereto contain only entries that are either 0 or positive facilitating their interpretation. Moreover, as $X$ contains inherently non-negative data (measurements of interaction with course content are either zero or positive for all learners), the non-negativity constraint also ensures, that its approximation $ P A^\top$ does not encounter any negative values.

Note, that  (\ref{xnmf}) presents an approximation of the observed learning behavior measurements. On one hand, some behavior (e.g. content queries outside of course system) can not be observed.
On the other hand, not all quantified actions can be represented perfectly by the learning styles, since also other individual factors, such as personal time constraints,  impact the observed learning behavior.

The number of columns $K$ in $P$ and $A$ corresponds to the number of common learning patterns that can be detected given the particular course data.  In further I assume that $K$ is known or its value can be reasonably specified by the researcher. For example, it can be determined using the number of different learning styles in a learning style model as reference.  Ideally, all learning styles of the assumed learning style system materialize in one of the common learning patterns. However, sometimes the related stylized behavior is not captured in the data and the corresponding learning patterns will not show up in the data, and thus can not be extracted. The choice of $K$ rests, therefore, on the learning styles system and on the available data.

The main objection is to compute the unknown part of (\ref{xnmf}) from the data given $K$, that is:
\begin{itemize}
\item common hidden learning patterns based on observable aspects of learning behavior, which are the columns of the hidden matrix $P$,
\item individual weights giving the personal affinity towards  each learning pattern, contained in the columns of the hidden matrix $A$,
\end{itemize}
such that both $P$ and $A$ are non-negative and the approximation is best in squared error sense:
\begin{align}
\min_{P\in\mathbb R_+^{p\times K}, A\in\mathbb R_+^{n\times K}} ||X - PA^\top||_F^2,\label{minp}
\end{align}
where $||M||_F^2$ denotes the squared Frobenius norm of a matrix $M$ and is equal to the sum of its squared elements.
The minimization problem in (\ref{minp}) corresponds to the non-negative matrix factorization (NMF) of $X$. This technique originally proposed by \cite{nmf} is used here to model the learning patterns and the respective affinities. 
{\small
\begin{algorithm}[H]\label{annls}
	\vspace{0.5cm}\hrule\vspace{0.5cm}
	\KwData{data matrix $X\in\mathbb R_+^{p\times n}$}
	\KwIn{the number of components $ K\in\mathbb N, K\leq p.$}
	\KwResult{$P\in\mathbb R_+^{p\times K}$ and $A\in\mathbb R_+^{n\times K}$}
	Initialize $P\in\mathbb  R_+^{p\times K}$\;
	\Repeat{convergence}{ 
		$A\leftarrow NNLS(P^\top, X^\top)$\;
		$P\leftarrow NNLS(A,X)$.
	}
	\Return{$P, A$}
\vspace{0.2cm}\hrule\vspace{0.2cm}
	\caption{\scriptsize NMF via alternating non-negative least squares.}
\end{algorithm}
}
The alternating non-negative least squares (NNLS) algorithm ( \cite{Gillis2012AMA}) is used to obtain the actual decomposition in $P$ and $A$ given $K$ and is presented in Algorithm \ref{annls}.
The $NNLS()$-Routine can be found in \cite{nnls}, p. 161, and is available as{ \it nnls} ~function in package {\it nnls} (\cite{pnnls}) in  R~ (\cite{rr}).

The decomposition resulting from Algorithm \ref{annls} is identifiable up to a scale factor. It can be rescaled without changing the overall approximation as:
\begin{align}
PA^\top = (PS)(S^{-1}A^\top)=P_SA_S^\top,\label{sc}
\end{align}
using a diagonal matrix $S=diag(s_1,s_2,\ldots,s_K)$ with $s_k,$  $k=1,\ldots, K$ as scale factors. For example, after obtaining the decomposition in $K$ learning patterns in $P$ and the respective affinities in $A$, one can scale the entries of $A$ to $[0,1]$ for each pattern on order to make them comparable.

With $P$ and $A$ obtained for a particular $K<p$ the model can be interpreted as a dimension reduction for the columns of $X$. Its $p$-dimensional columns (learners) can be represented by their $K$-dimensional individual affinities to the respective learning patterns. For instance, consider a  learner with a behavioral pattern as in Figure (\ref{pic}) (learner \#1)  and with affinities $(0.5, 0.25)$ regarding pattern 1 and pattern 2 respectively, each consisting of {\it five} variables. If there is another learner (learner \#2) exhibiting affinities $(0.1,0.8)$ regarding the same patterns, then both learners can be compared by just looking on their {\it two}-dimensional affinities \footnote{In this case, learner \#1 shows higher inclination to the first pattern, whereas learner \#2 prefers the second pattern.}. The comparison becomes more straightforward as when using the original data of five variables for each learner. 

 In general, instead of considering all $p$ measurements of learners' behavior, one can use the corresponding $K$ affinities for exploring the data e.g. for clustering and visualization. Let $M_{i,j}$ denote the $ij$th element and $M_{\cdot,j}$ the $j$th column of matrix $M$. Then, the observed behavior of learner $j$ in the data matrix $X$ can now be approximated as:
\begin{align}
X_{\cdot,j} \approx P A_{\cdot,j} = \sum_{i=1}^{K} A_{i,j}P_{\cdot,j}.\label{xx}
\end{align}
The distribution of the affinities among learners provides new insights on prevailing learning patterns and, given some additional information as  the final performance of the learners, can be used to determine the role of the affinities therein. I explore this fact further in the results section.

\section{Results and  Discussion}
In this section, the proposed model is applied to actual course interaction data. I describe the data and present the resulting decomposition in common learning patterns and corresponding affinities. Finally, a test strategy for checking the learning style balance of the course is proposed and the results are discussed.
\subsection{Course interaction data}
The analysis is based on the data of 111 students of a statistics course during one academic semester.  The course was a follow-up to a prior basic course and took place completely online due pandemic contact restrictions.  The course interaction data contains only the measurements of the learners who actually completed the course and took the final exam. 
The structure of the course includes ten topics each scheduled to be completed by a particular due date. There were all together about 180 materials available. In particular, each topic included lecture summaries, videos with lectures and examples, links to online resources, practical exercises, and quizzes for each topic.
\begin{table}[H]
{\scriptsize
\begin{tabular}{llll}
name&meaning&addressed style\\\hline

a\_try&average number of attempts on quizzes&  active, sensing\\
a\_tryx&average attempts on quizzes after the 50\% threshold&  active, achieving\\
opt\_try& average attempts on optional final quizzes& active, global\\
a\_time&average time passed when solving quiz&reflective, sensing\\
a\_score& average score on quizzes&sensing, achieving\\
th\_score&scores on quiz questions with theoretical emphasis& intuitive\\
st\_score&scores on quiz questions with standard tasks& sensing\\
m\_score& average score exceeding the 50\% threshold&achieving\\
p\_noans& mean number of quiz questions without an answer&global, surface\\
q\_dur& queries on the content during quiz solving&reflective, surface\\
%quizcompletion
p\_quizc&proportion of completed quizzes&active, intuitive, sequential\\
%content
p\_cont&proportion of retrieved lecture notes& reflective, intuitive\\
p\_prac& proportion of retrieved exercises& reflective, sensing\\
p\_links& proportion of clicked links& reflective\\
p\_vids& proportion of seen videos&reflective\\
%review/skip
review\_q& number of quizzes reviewed later&active, global\\
%timely
a\_early&average time surplus of early completions&sequential\\ % 
on\_time&proportion of activities completed on time&sequential\\
a\_late&average delay of late completions&global\\
%forum
words\_frm& total number of words in forum posts&active\\
posts\_frm& total number of posted messages&active\\
\hline
\end{tabular}
}
\caption{\scriptsize List of the 21 features computed  from the course interaction data with the intended style identification.}\label{feat}
\end{table}
The questions in quizzes were drawn randomly from a larger question pool or filled with random numbers. The number of visits to the quizzes was not limited: students could repeat their attempts as often as they wish.
 The incorporated Q\&A Forum served as a communication tool. 
Besides the materials available in the online course system, synchronized online lectures took place, for which no attendance information is available. 

Based on the stylized data patterns for each learning style and the recommendations regarding their identification given in \cite{graf2}, I compute 21 features from the course interaction data  (Table \ref{feat}) to reflect the stylized behavior patterns in quizzes, content retrieval, task completion, and other aspects of stylized behavior of the mentioned learning styles.
 As seen from Table \ref{feat},  learning behavior concerning the input dimension (verbal and visual learning styles) can not be quantified from the data. The provided course materials are rather mixed, containing both visual and verbal content. Therefore, the corresponding stylized behavior patterns are not reflected in the gathered course interaction data. Under this consideration, the input dimension is excluded from further analysis. Based on the data from the course and the computed variables of learning behavior in Table \ref{feat}, I should be able to identify the common learning patterns connected to the learning styles: active / reflective, sensing / intuitive, sequential / global, and surface / achieving.

\subsection{Decomposition of the data in $P$ and $A$}
To find $P$ and $A$ of the approximation in (\ref{xnmf}) using Algorithm \ref{annls}, one has to specify $K$, the number of common learning patterns supposedly prevalent in the data. I choose $K=8$ based on the eight learning styles, which stylized behavior I capture with the measurements in Table \ref{feat}: active, reflective, sensing, intuitive, global, sequential, surface, and achieving. 

Prior to starting the optimization, I scale the $(21\times 111)$ data matrix by dividing the rows by their maxima to  bring the variables on the same scale [0,1]. The scaled data  (denoted as $X$) becomes then the  matrix to approximate with the product of $P$ and $A^\top$.
Applying the algorithm for non-negative matrix factorization to $X$ returns the actual decomposition. The extracted $K$ learning patterns are contained in the resulting matrix $P$ and the corresponding individual affinities are contained in $A$.
 The entries of $A$ are then rescaled to be between 0 and 1 by left-multiplying it with $S^{-1}$, where $S$ is diagonal containing the row means of $A$ on the diagonal. The affinities towards each of the patterns are then on the same scale. $P$ have to be now  right-multiplied with $S$, so that the overall approximation does not change, as given in (\ref{sc}).

\begin{figure}[H]
	\centering
	\includegraphics[scale=0.6]{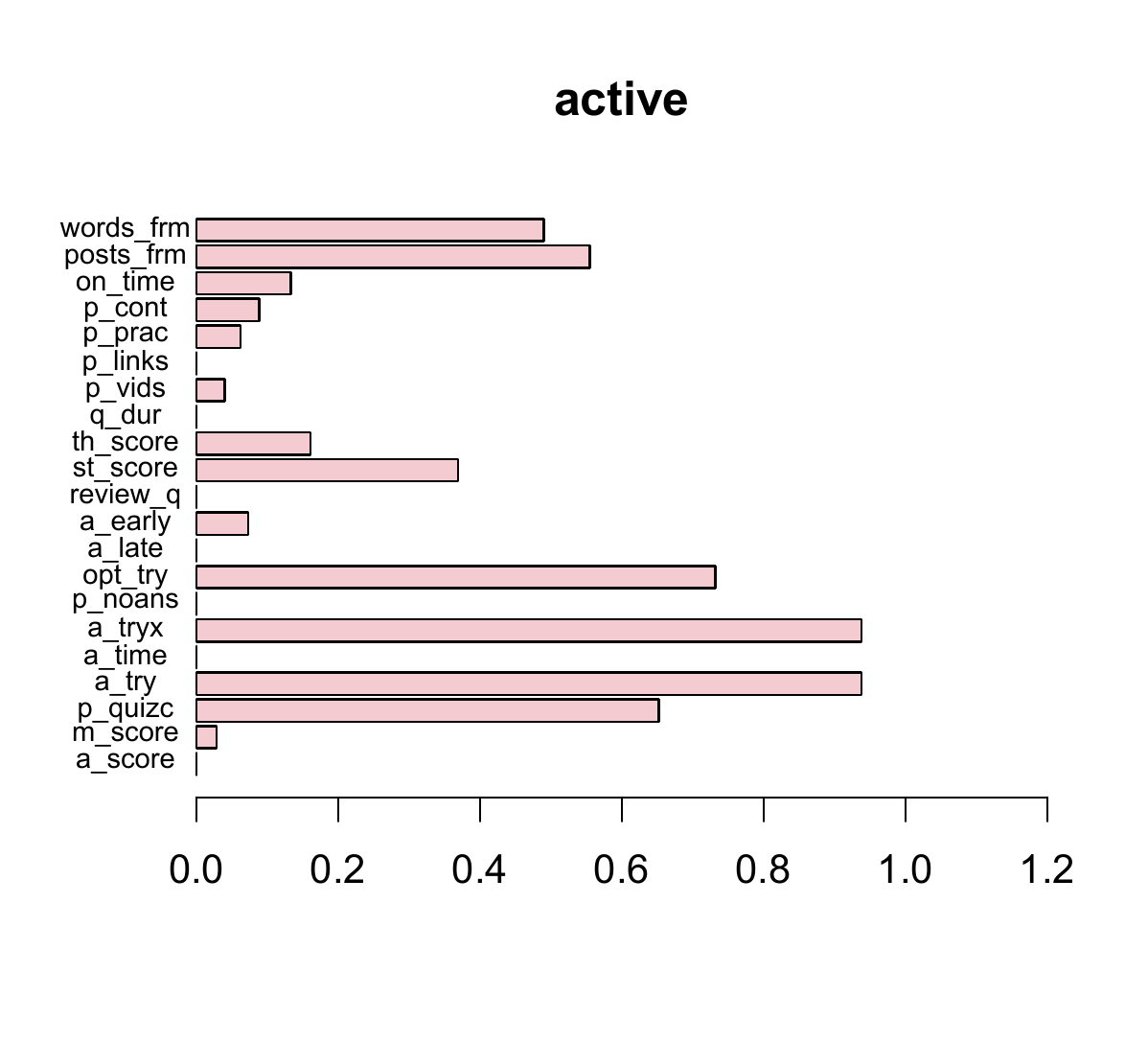}\includegraphics[scale=0.6]{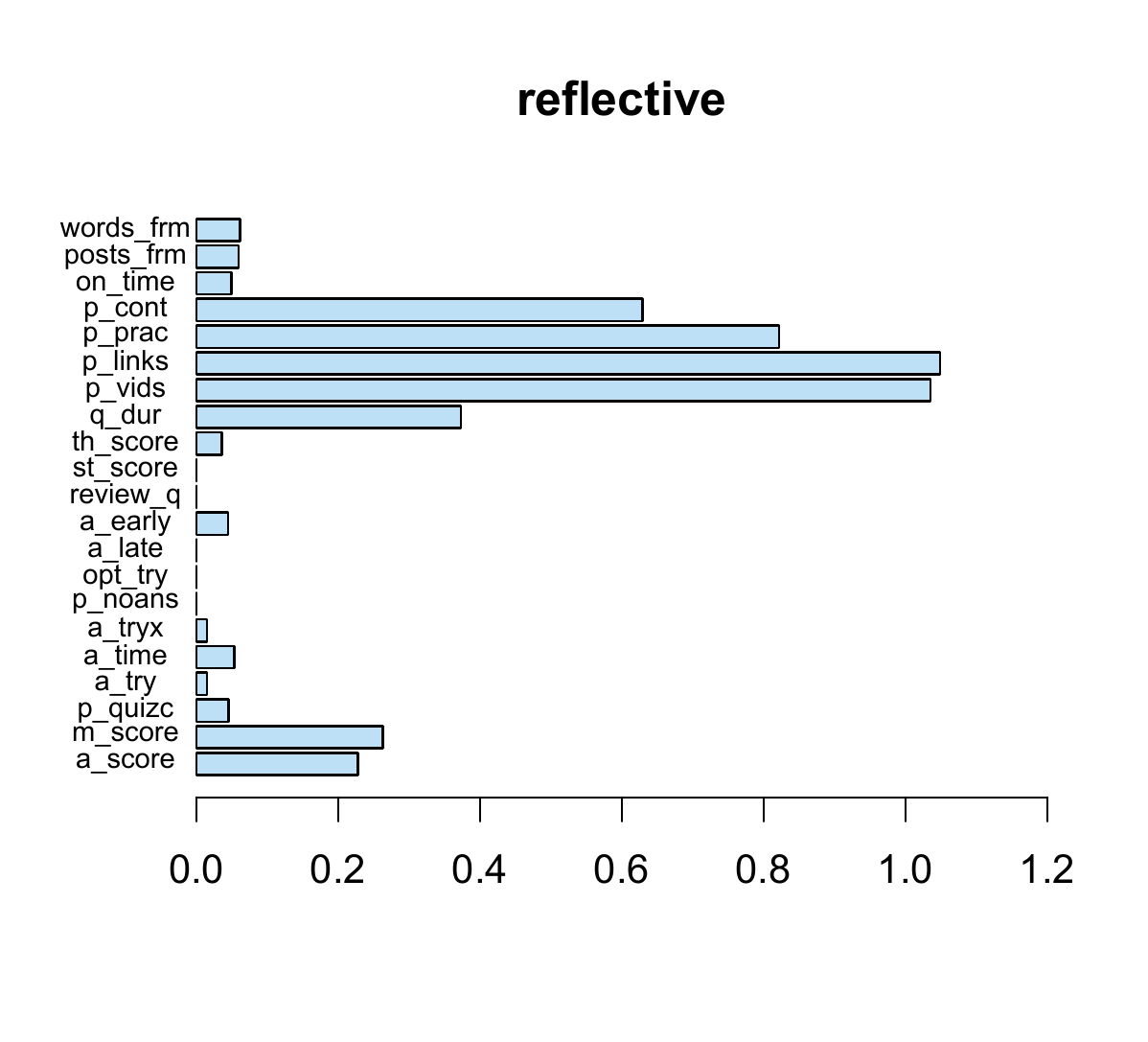}
	\includegraphics[scale=0.6]{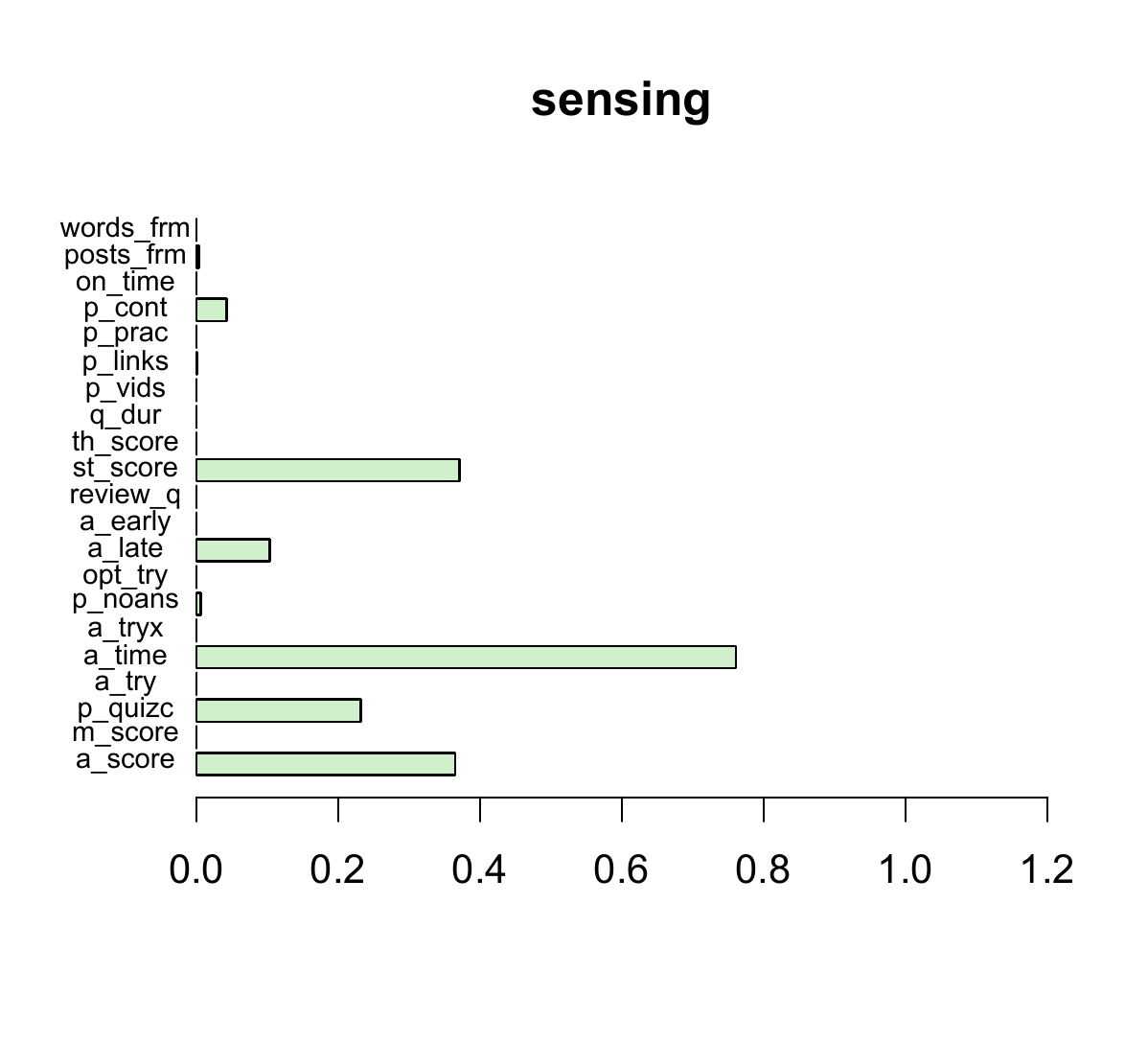}\includegraphics[scale=0.6]{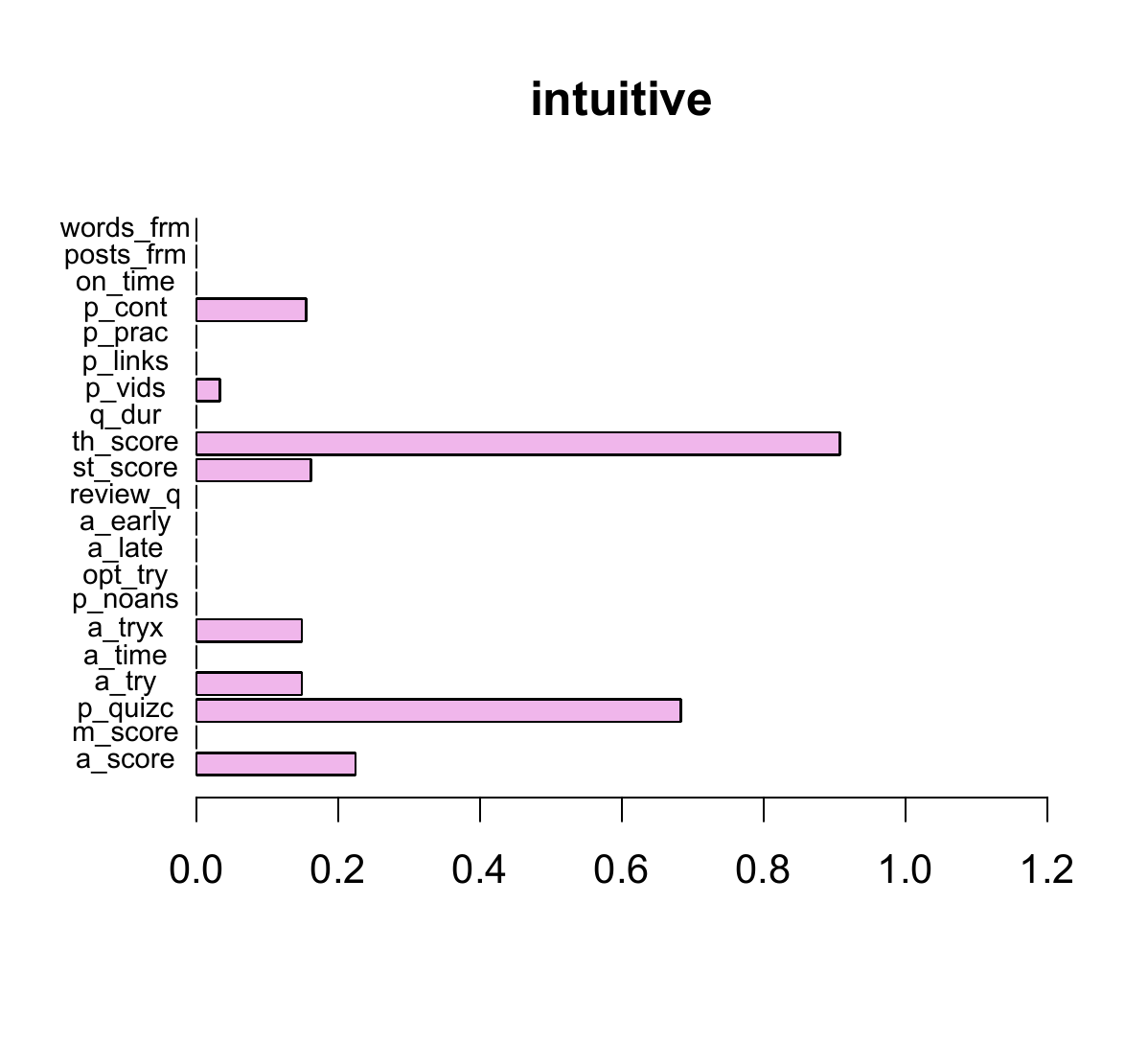}
	\caption{\scriptsize{The extracted leaning patterns related to the learning styles active / reflective and sensing / intuitive. The bars' lengths correspond to the coefficients (horizontal axis) of the respective features (vertical axis).
			}}\label{comps_bars1}
\end{figure}

\begin{figure}[H]
	\centering
	\includegraphics[scale=0.6]{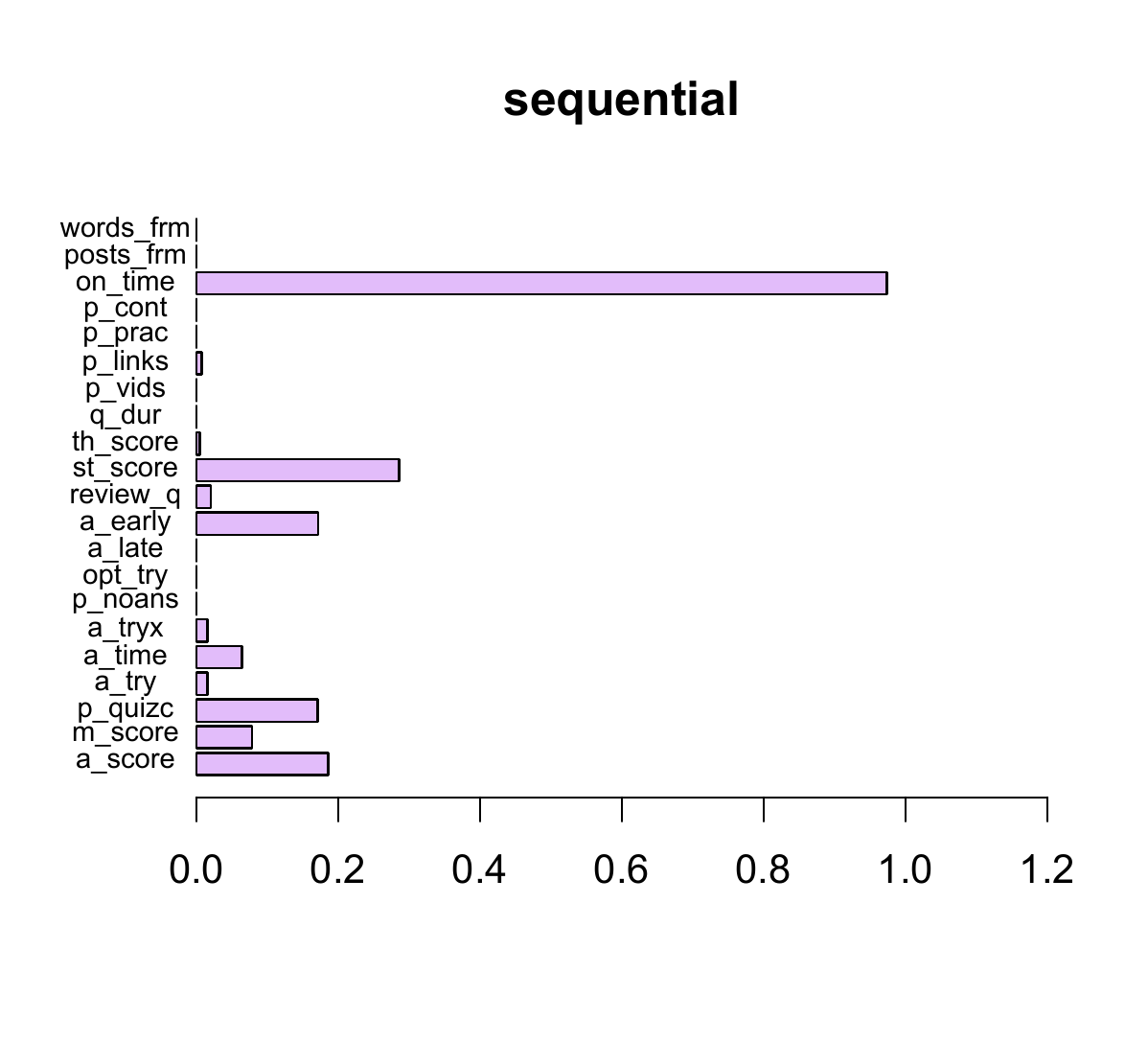}\includegraphics[scale=0.6]{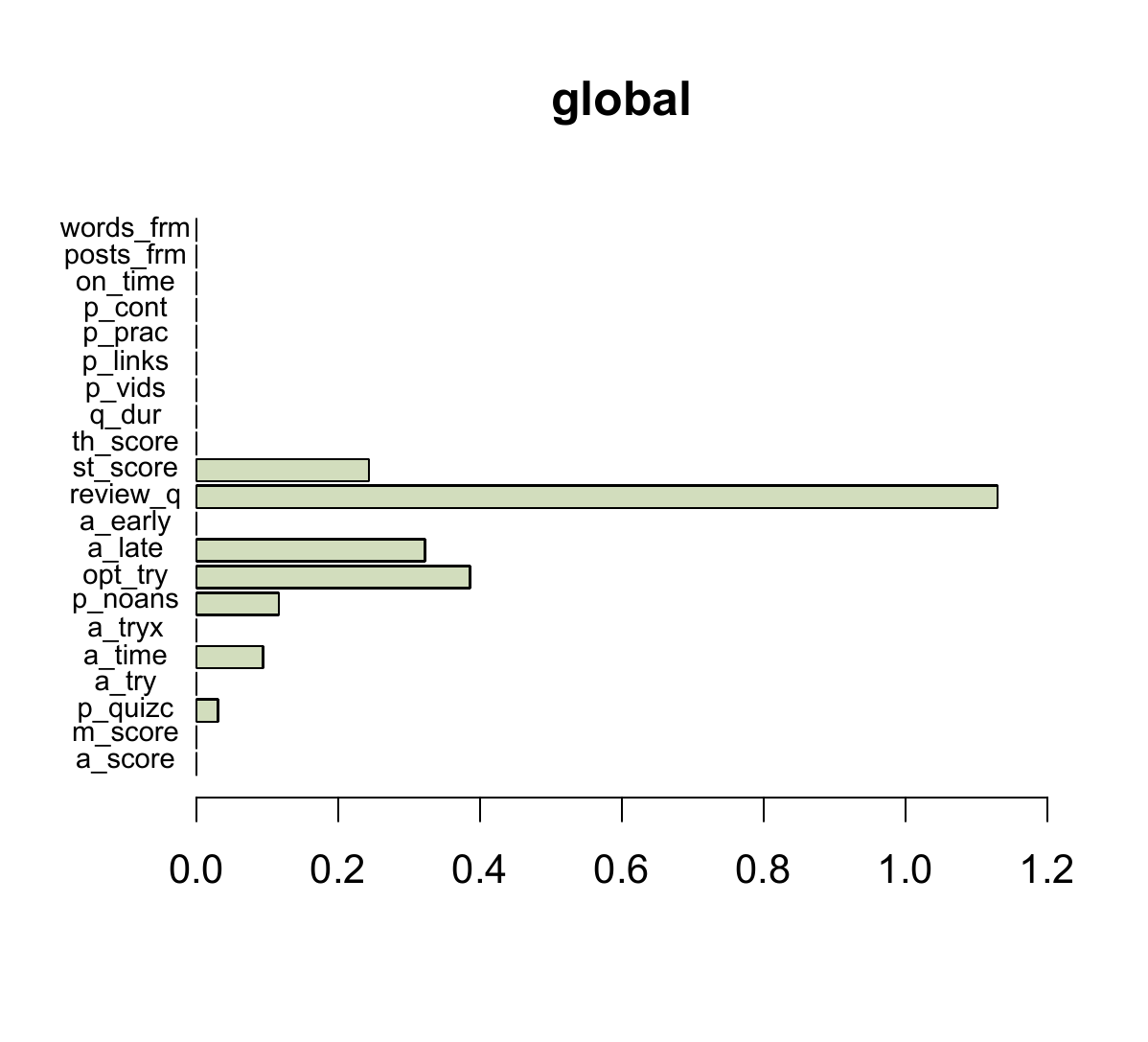}
	\includegraphics[scale=0.6]{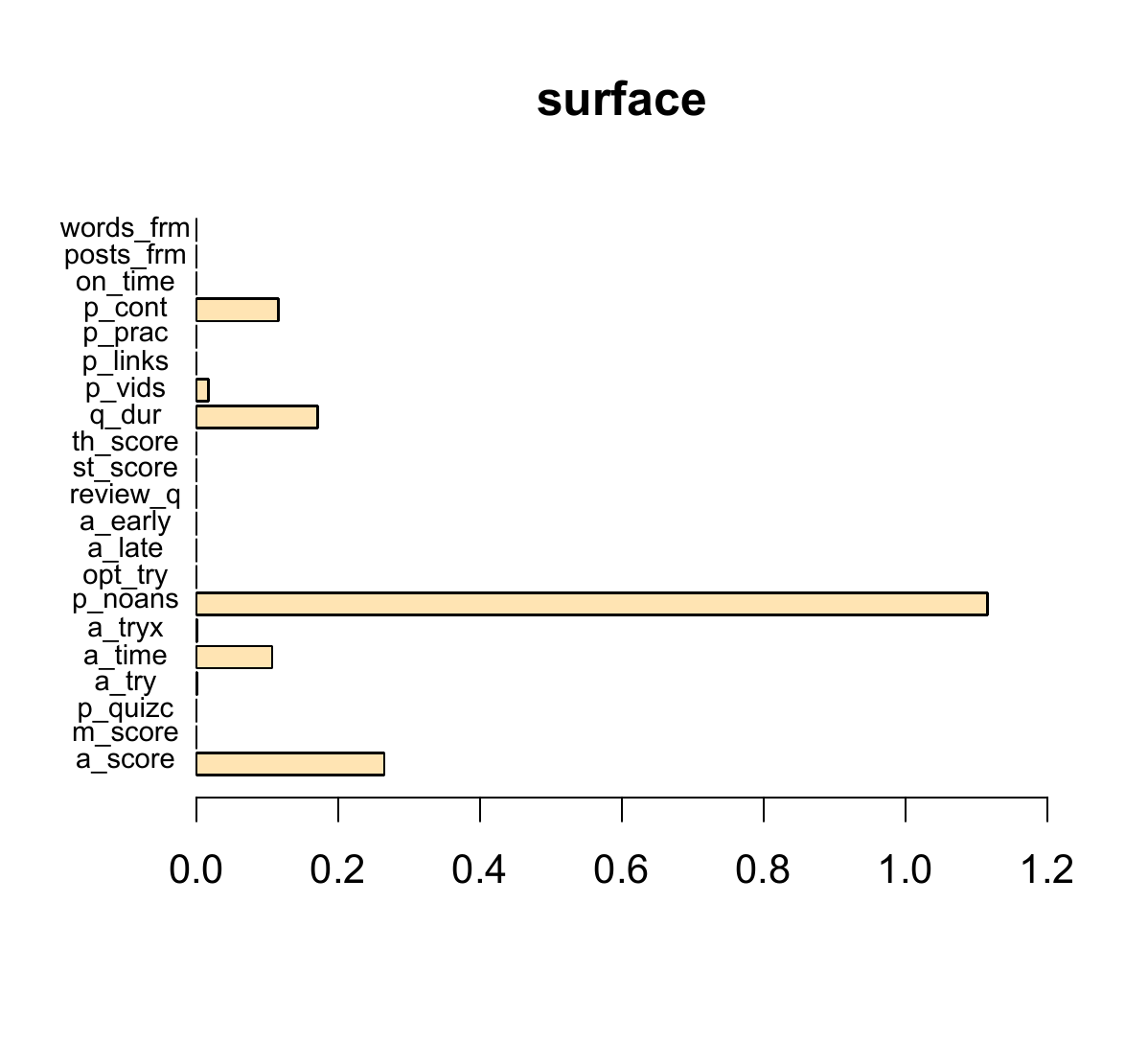}\includegraphics[scale=0.6]{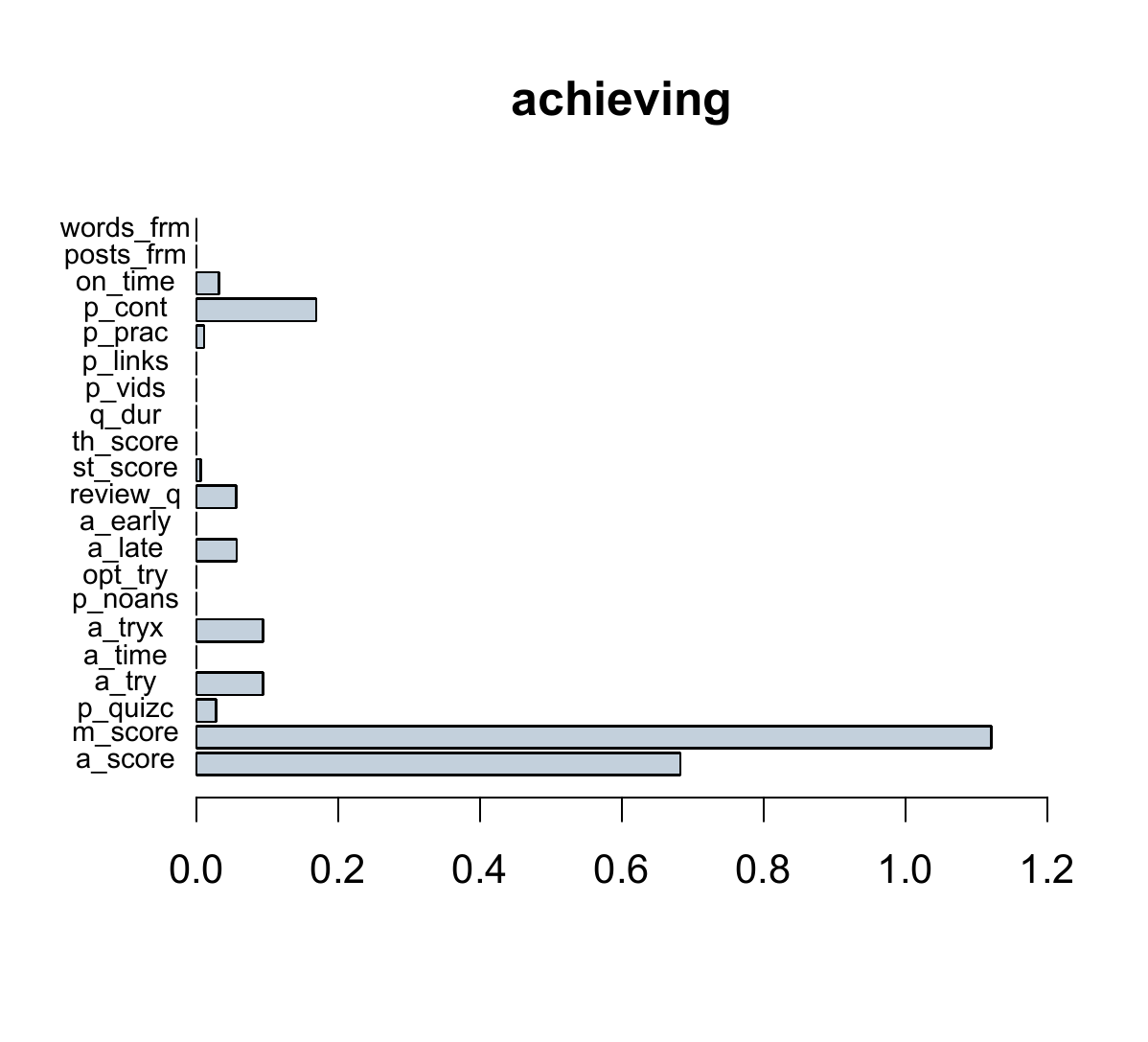}
	\caption{\scriptsize{The extracted leaning patterns related to the learning styles sequential / global and surface / achieving. The bars' lengths correspond to the coefficients (horizontal axis) of the respective features (vertical axis).
			}}\label{comps_bars2}
\end{figure}

To relate the obtained patterns in $P$ to the eight learning styles above, let's inspect the coefficients in the columns of $P$. Significant coefficient magnitudes, distinguishing particular patterns from the others, indicate, combined with the entries of Table \ref{feat}, possible links to the adopted learning style system. The extracted patterns are presented in Figures \ref{comps_bars1} and \ref{comps_bars2}, where they are labeled according to the stylized behavior expressed by the coefficient magnitudes.

The first and the second patterns must correspond to active / reflective styles as they discriminate between such action loaded behavior as solving quizzes, doing optional exercises, posting comments in forum (active, red) and isolated learning by studying and overthinking the available materials in the reflection mode (reflective, blue).  The third and the fourth patterns seem to address the sensing versus intuitive learning behavior. The third pattern weights such measurements as time in quizzes, average scores and scores on standardized tasks the most (sensing, green), whereas the fourth emphasizes visits to lecture notes, the scores in theoretical questions and quiz completion rate suggesting an intuitive approach (pink).

The fifth pattern, labeled as sequential, gives significant weights to on-time or earlier submissions and completion rate (violet) as opposed to the global strategy, where reviewed materials, late submissions, and attempts on final quizzes are distinguished (olive). Finally, achieving learners glance with best scores producing the largest coefficients (gray) wheres the surface pattern (orange) gives the highest importance to the proportion of questions left without any answer by average quiz scores and in the presence of (first-time) queries to the materials while trying to solve the exercises, indicating shallow learning afford just enough to pass.
As seen from the Figures \ref{comps_bars1} and \ref{comps_bars2}, the extracted patterns are rather sparse: numerous measurements get zero-weighted imposing irrelevance to the particular pattern. Some other features get a positive weight with small magnitude suggesting they might be also dispensable. To get a more systematic understanding of the importance and stability of the patterns, I construct confidence intervals for their coefficients using non-parametric bootstrap (\cite{efron93}, \cite{bootPCA}) with $B=10^4$ replications\footnote{I assume here that the data is a representative sample from the corresponding population, which is crucial to ensure consistency of bootstrap results.}. The confidence intervals help to verify whether the coefficients, used to connect the patterns to the learning styles, significantly differ from zero. 
The resulting $99\%$ point-wise confidence intervals are depicted in the next Figures \ref{ci1} and \ref{ci2}.
\begin{figure}[H]
	\centering
	\includegraphics[scale=0.5]{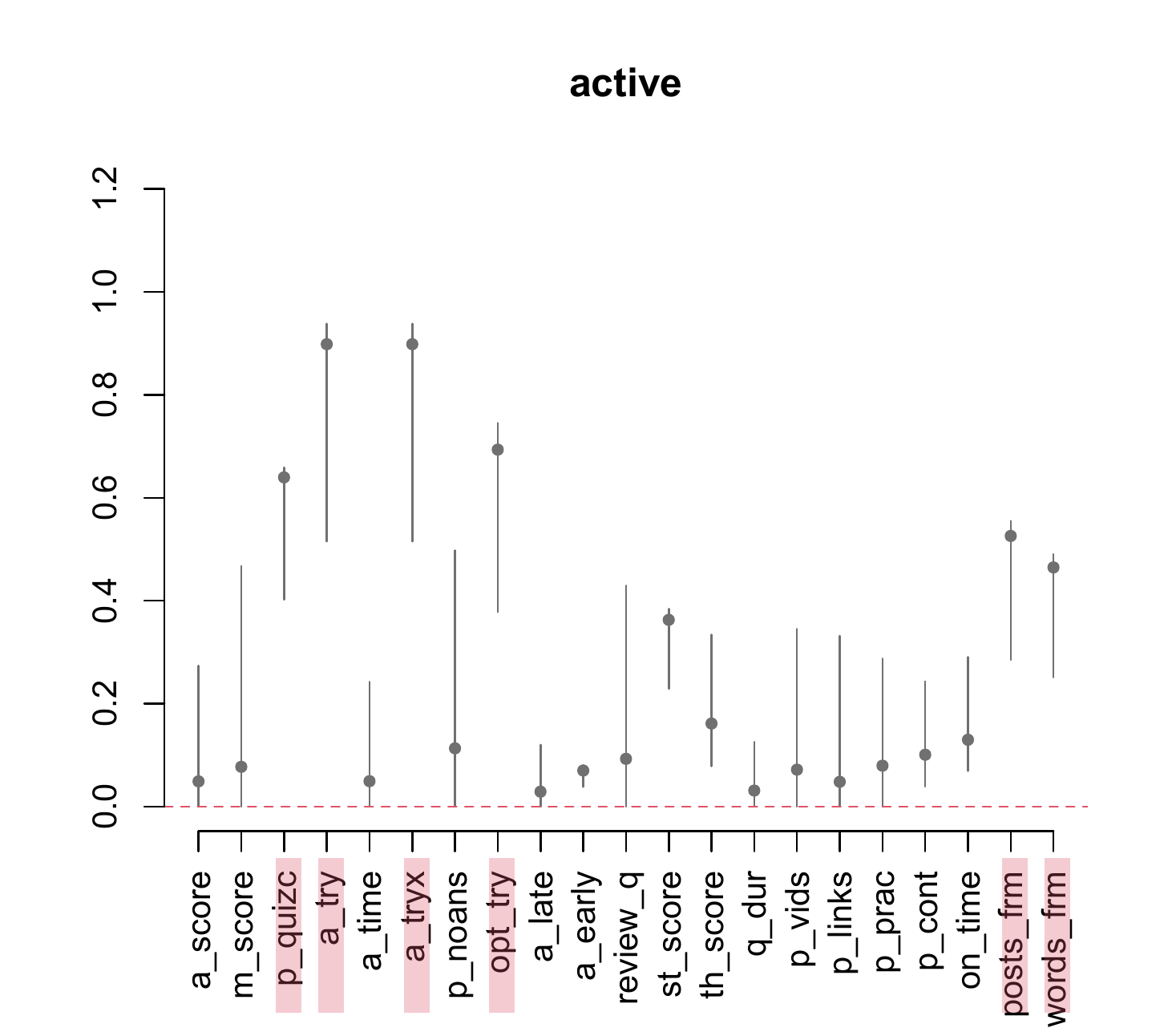}\includegraphics[scale=0.5]{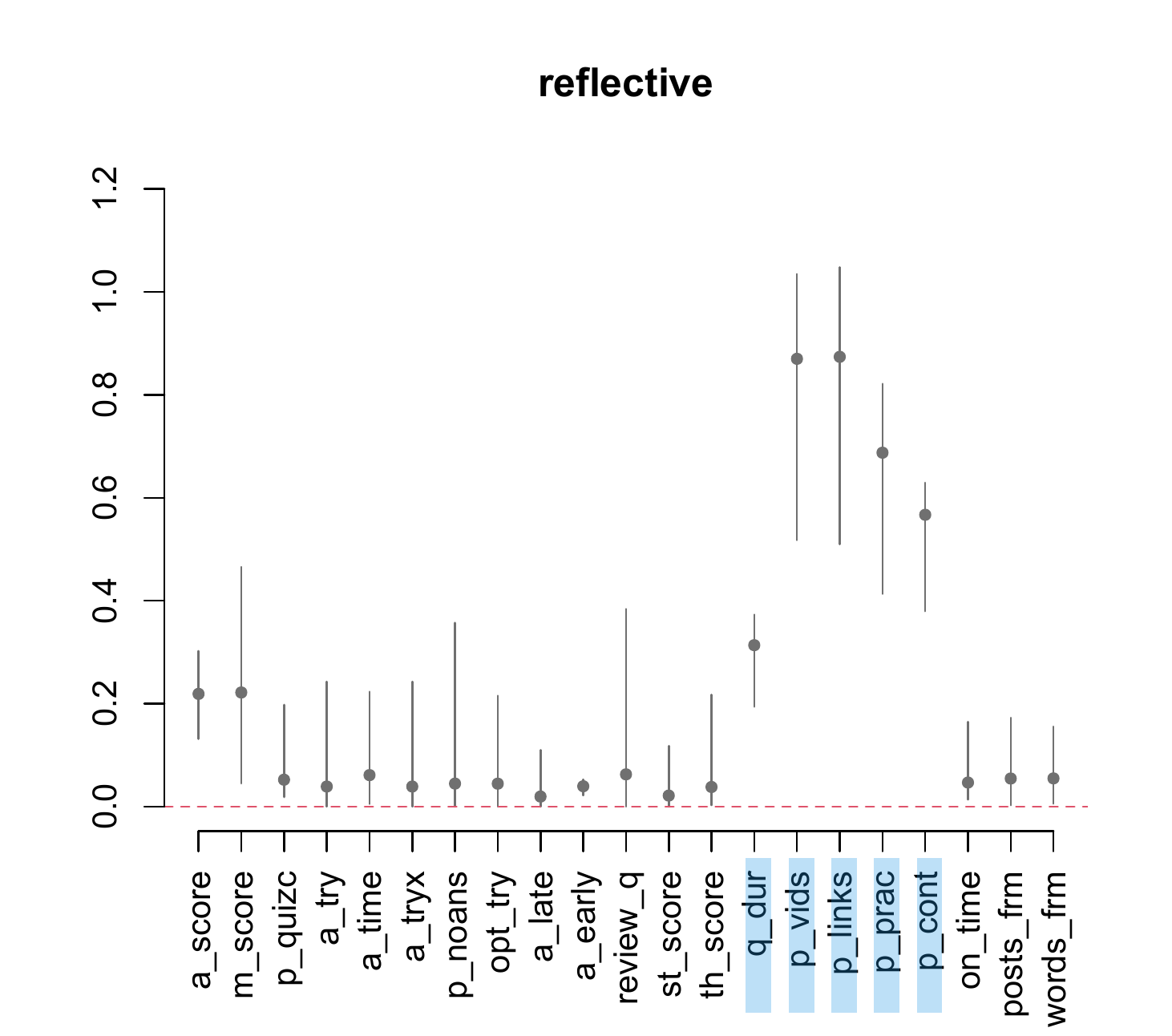}
	\includegraphics[scale=0.5]{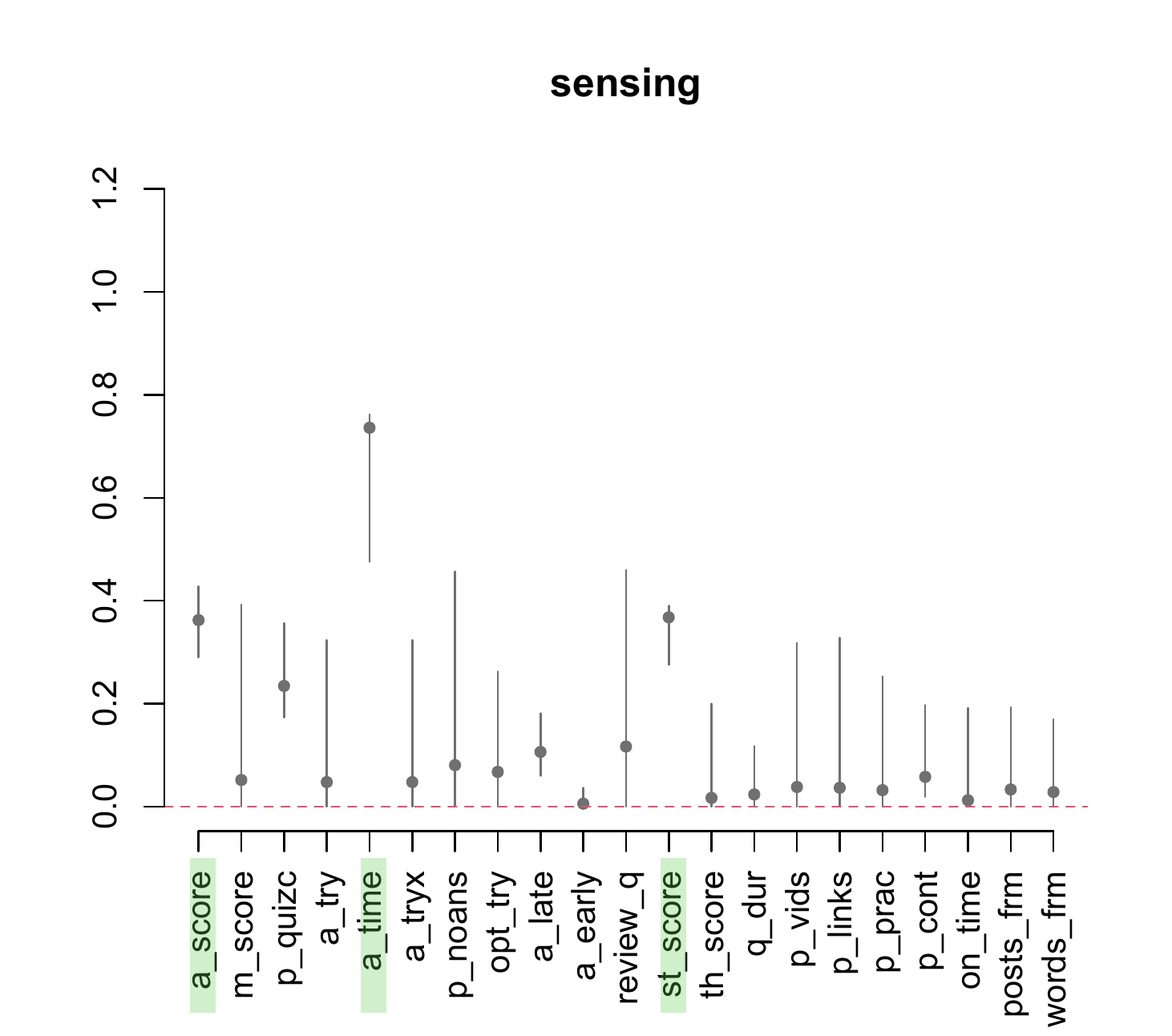}\includegraphics[scale=0.5]{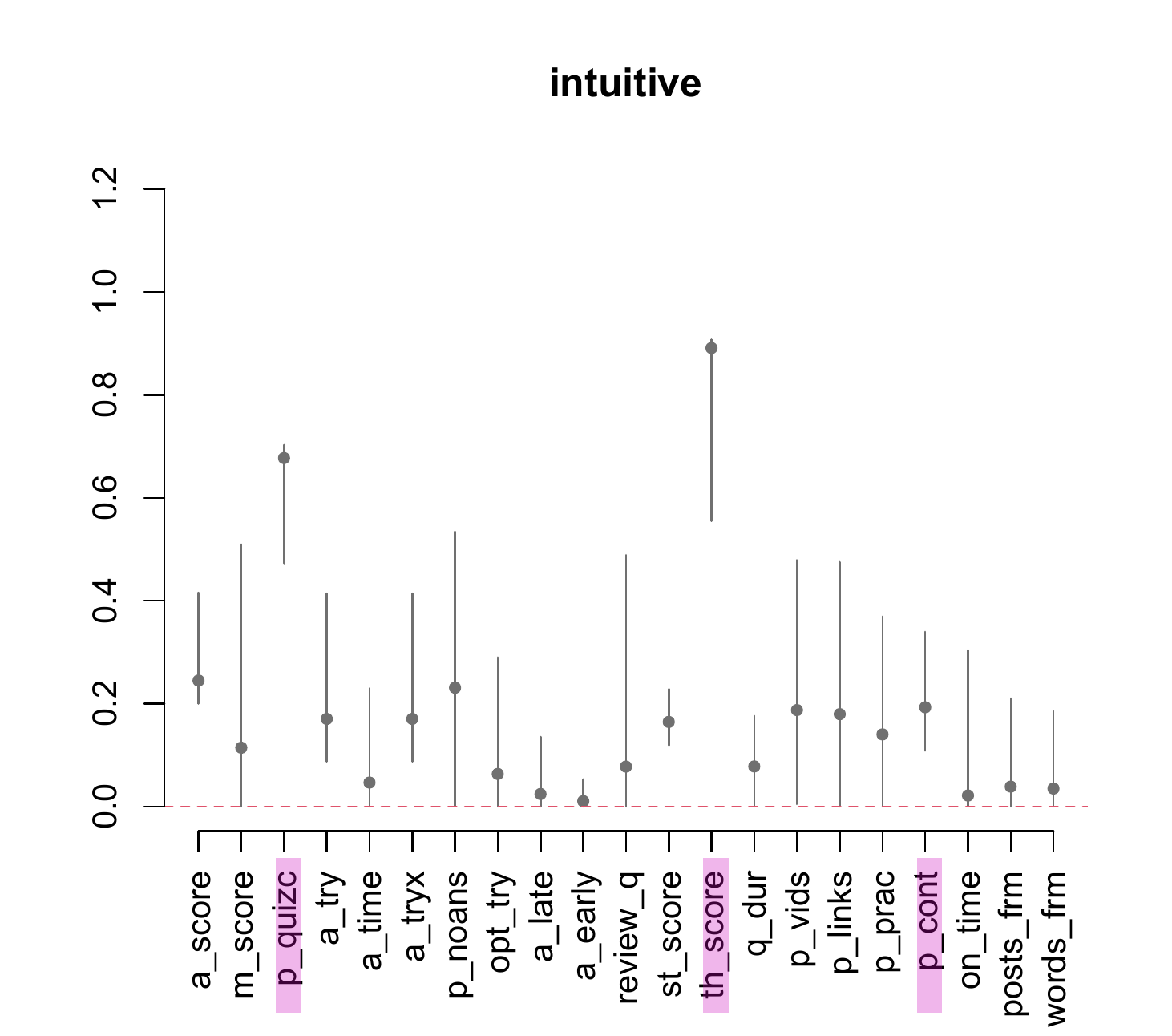}
	\caption{\scriptsize{Bootstrap means (circles) and 99\% confidence intervals (lines) for the learning patterns related to the learning styles active / reflective and sensing / intuitive. The features used for linking the patterns to the styles  are highlighted.
			}}\label{ci1}
\end{figure}
\vspace{-0.4cm}
Inspecting Figures \ref{ci1} and \ref{ci2} reveals that low coefficients often do not significantly differ from zero and may indicate features of low importance for the pattern. The high-weight "pattern defining features" used to connect the patterns to the learning styles (highlighted in the respective colors), are all significant. Thus, checking the confidence intervals ensures, that the linking of the extracted patterns to the learning styles is substantial.

\begin{figure}[H]
	\centering
	\includegraphics[scale=0.5]{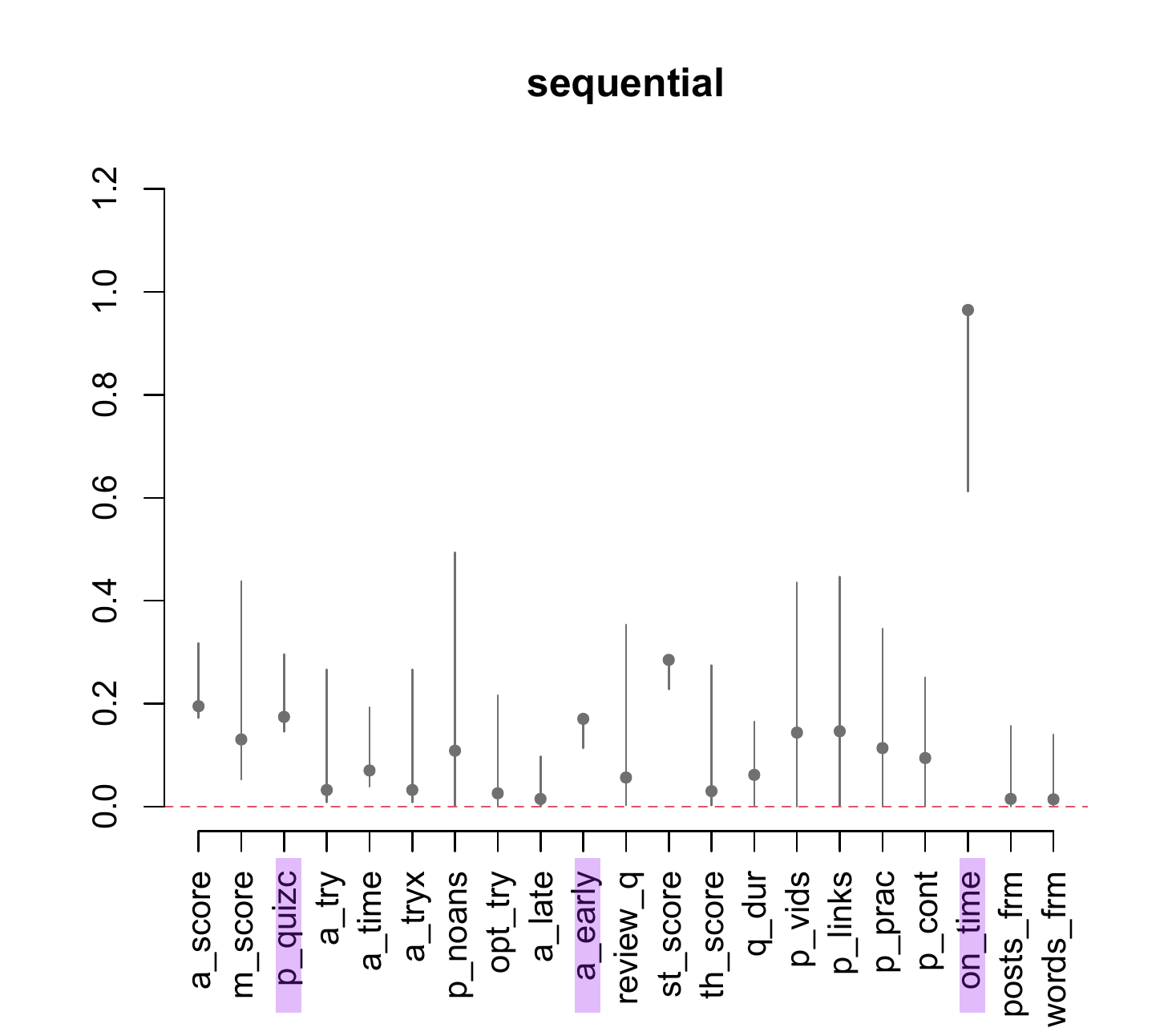}\includegraphics[scale=0.5]{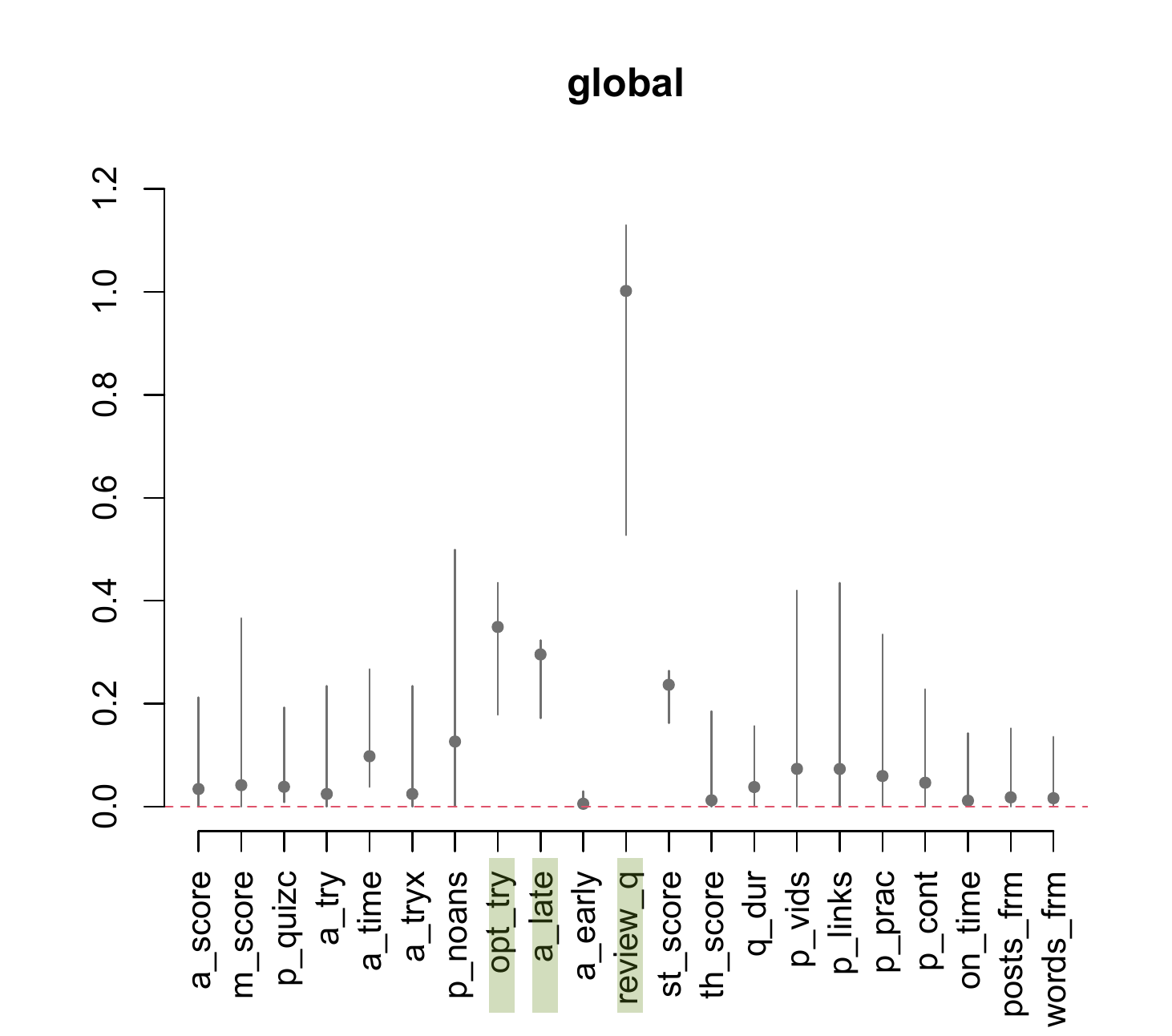}
	\includegraphics[scale=0.5]{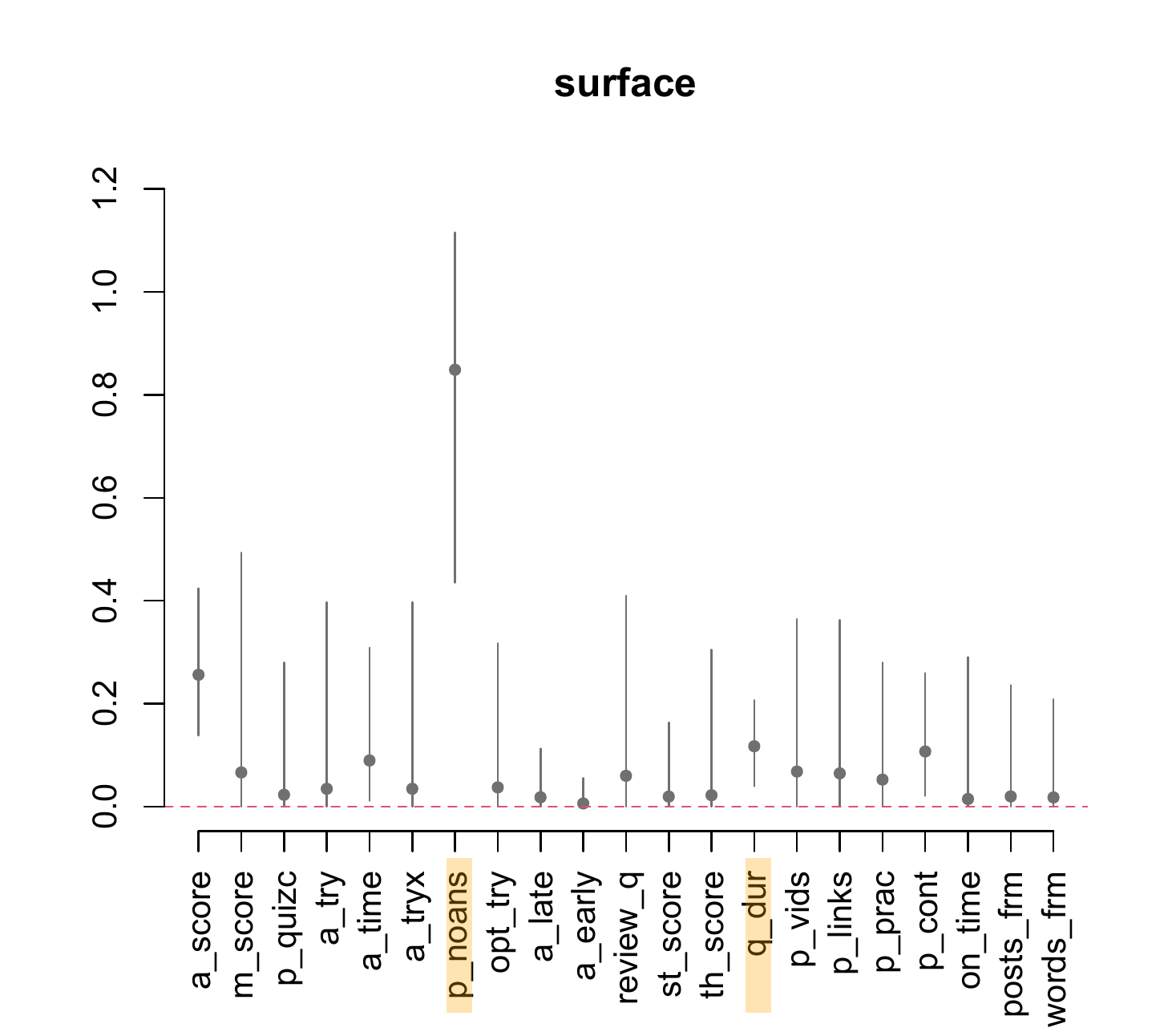}\includegraphics[scale=0.5]{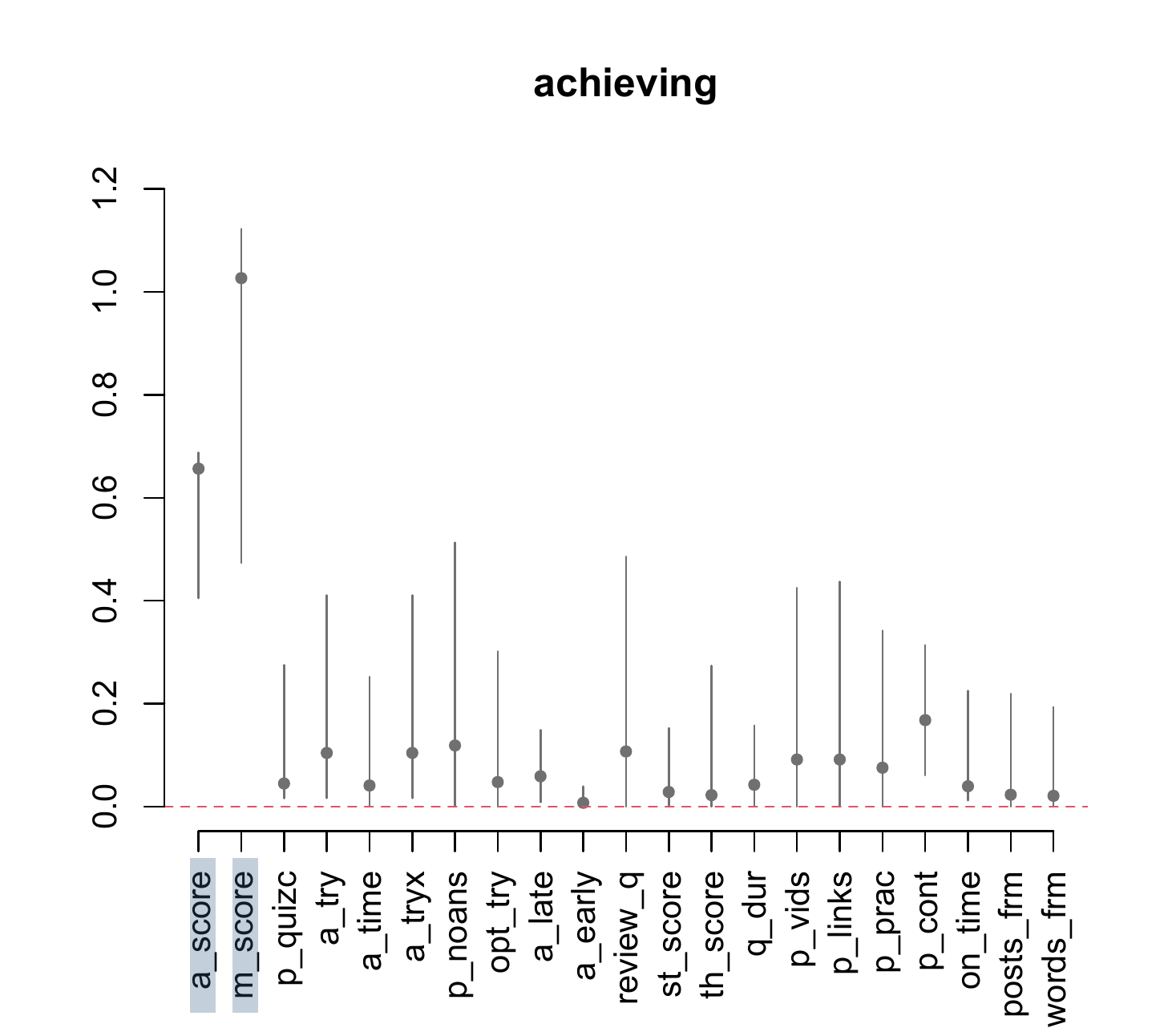}
	\caption{\scriptsize{Bootstrap means (circles) and 99\% confidence intervals (lines) for the learning patterns related to the learning styles sequential / global and surface / achieving. The features used for linking the patterns to the styles  are highlighted.
			}}\label{ci2}
\end{figure}
\begin{figure}[H]
	\centering
	\includegraphics[scale=0.6]{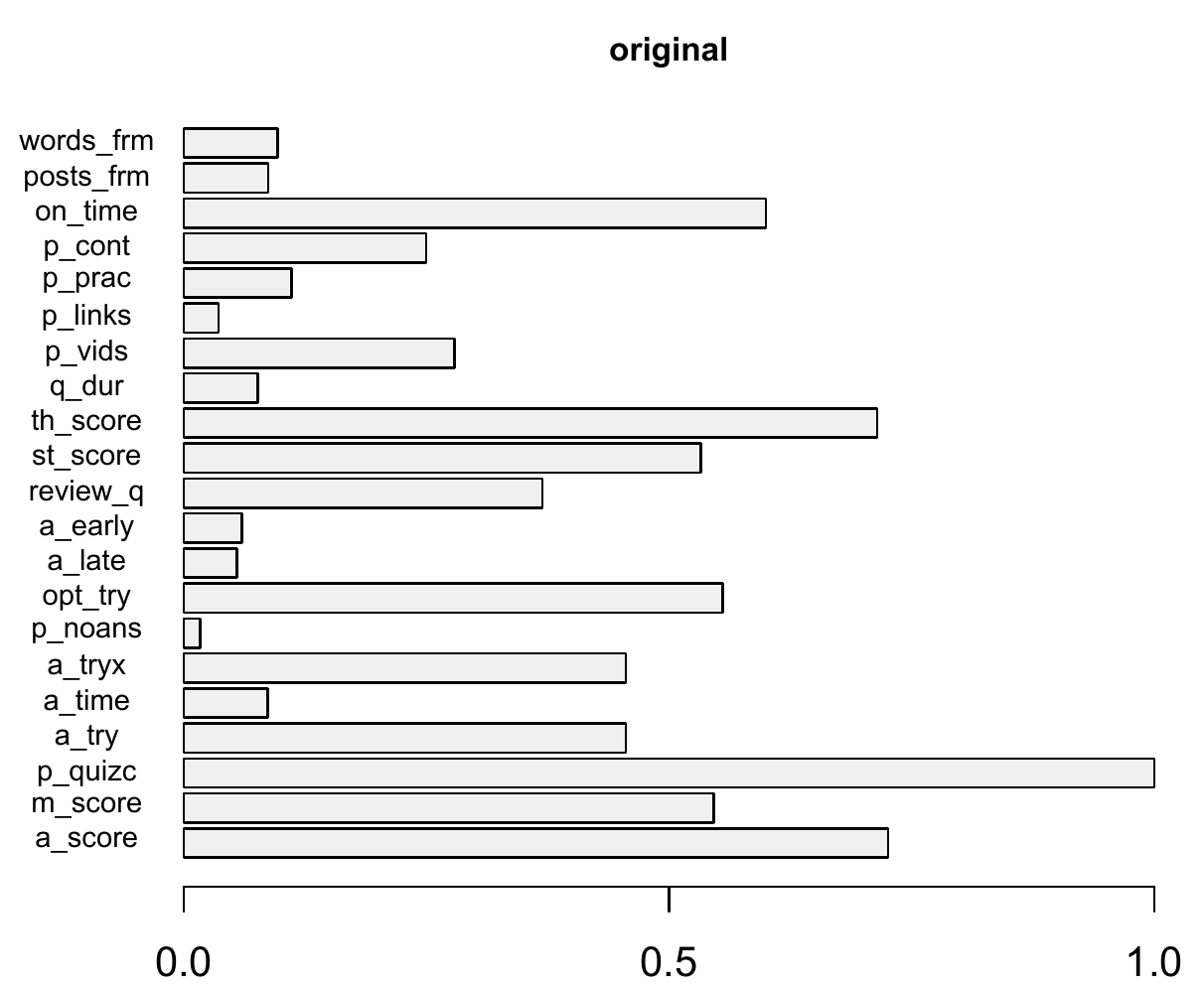} \includegraphics[scale=0.6]{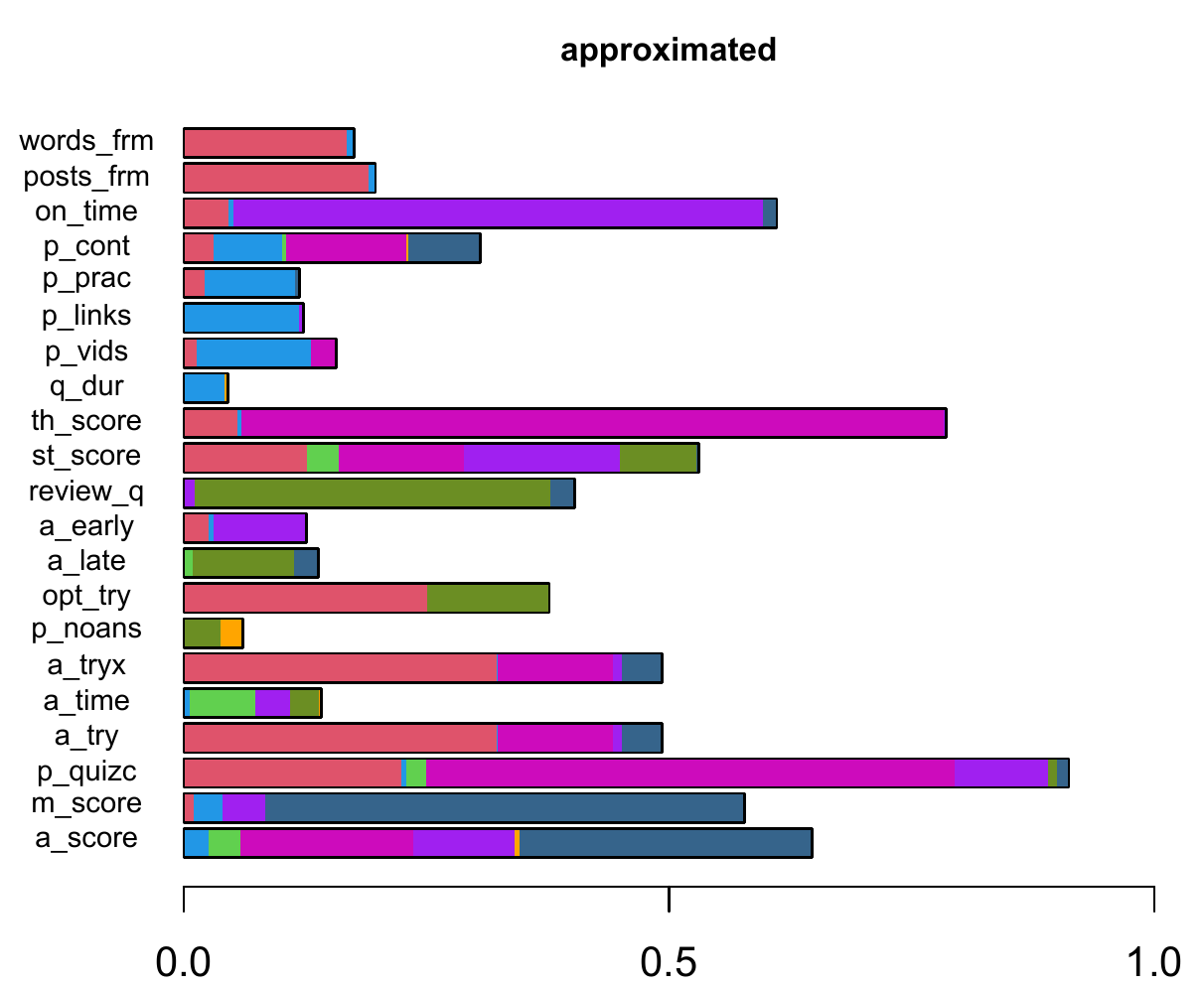} \\
\vspace{0.2cm}
\includegraphics[scale=0.33]{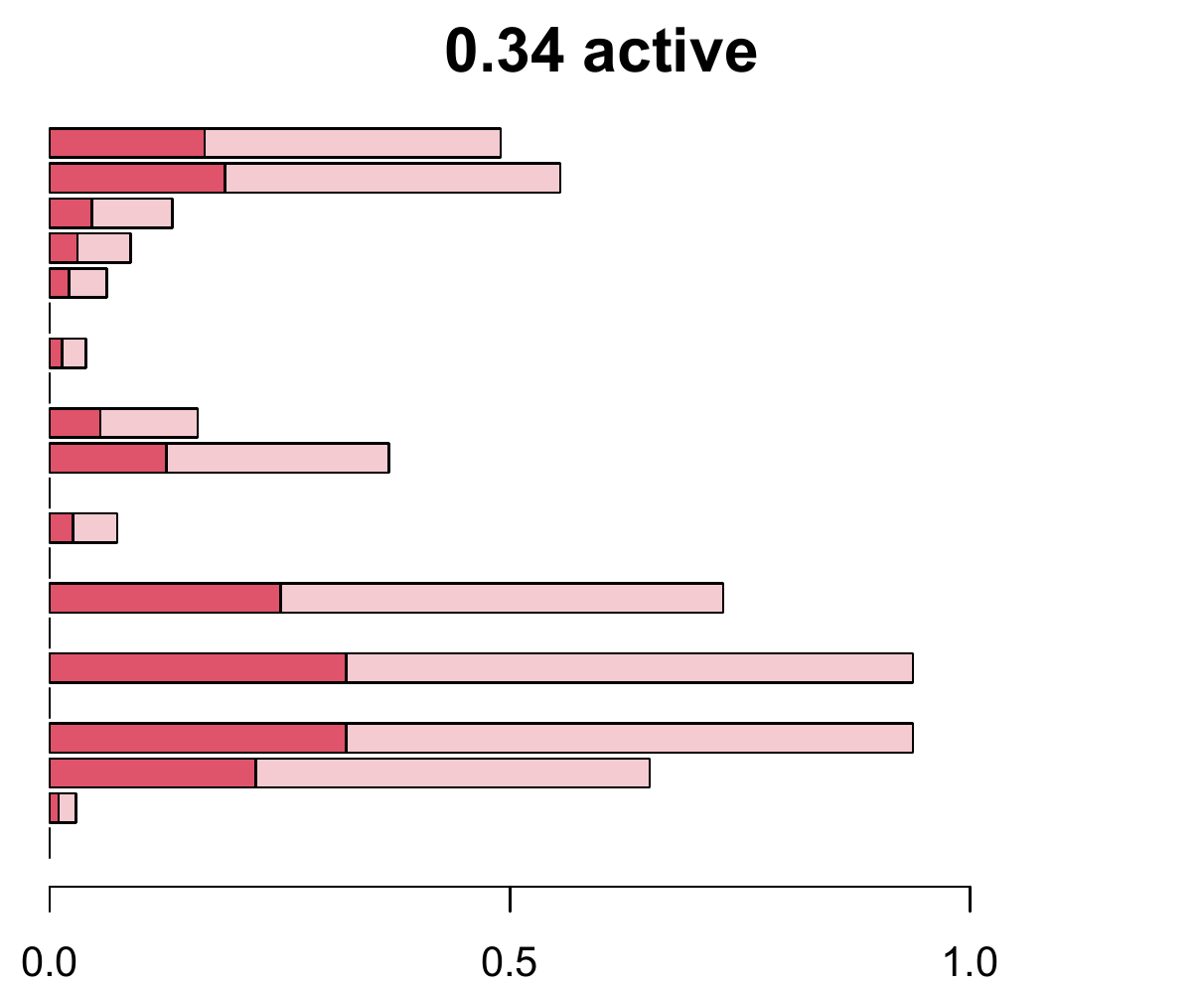}\includegraphics[scale=0.33]{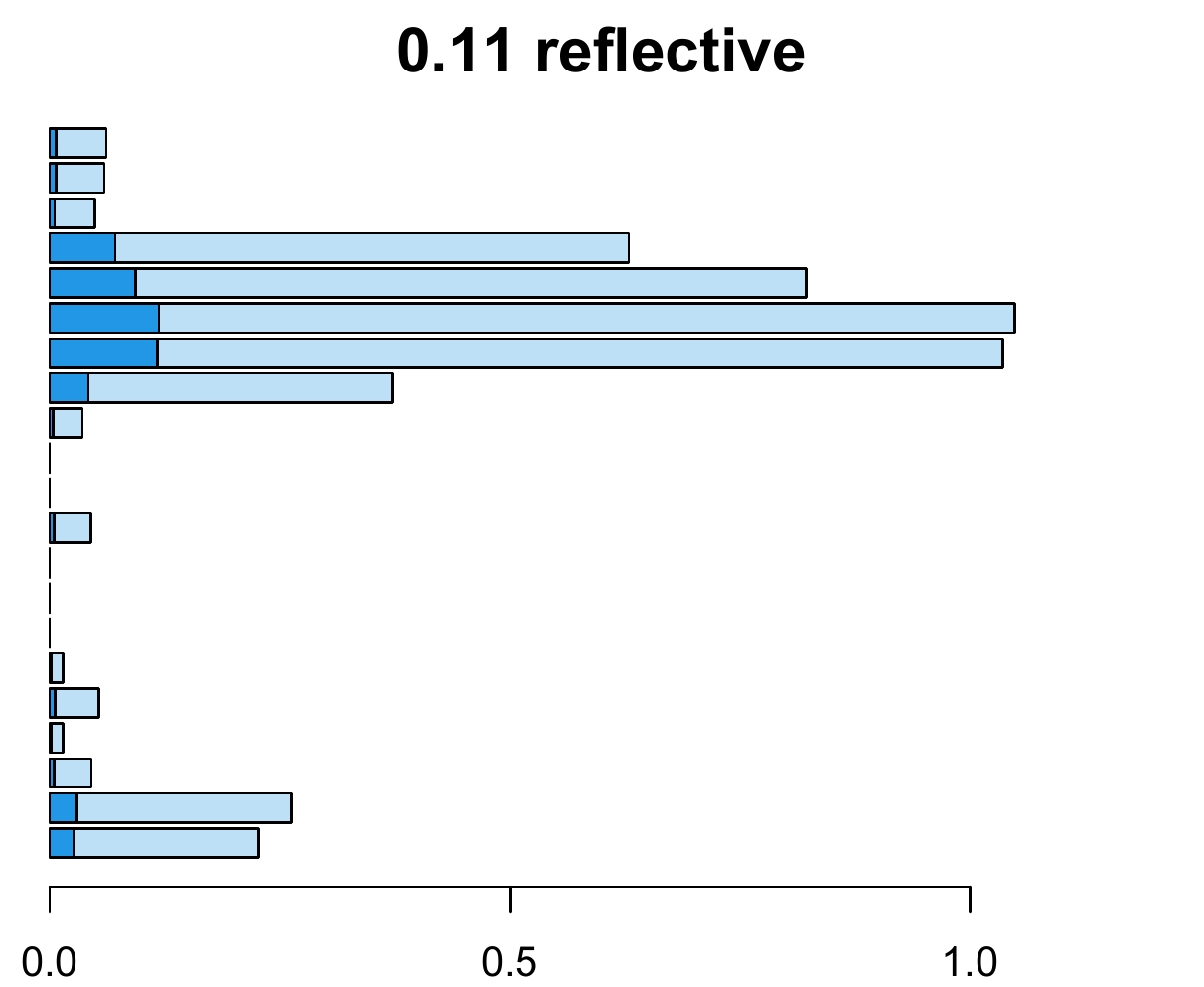}
\includegraphics[scale=0.33]{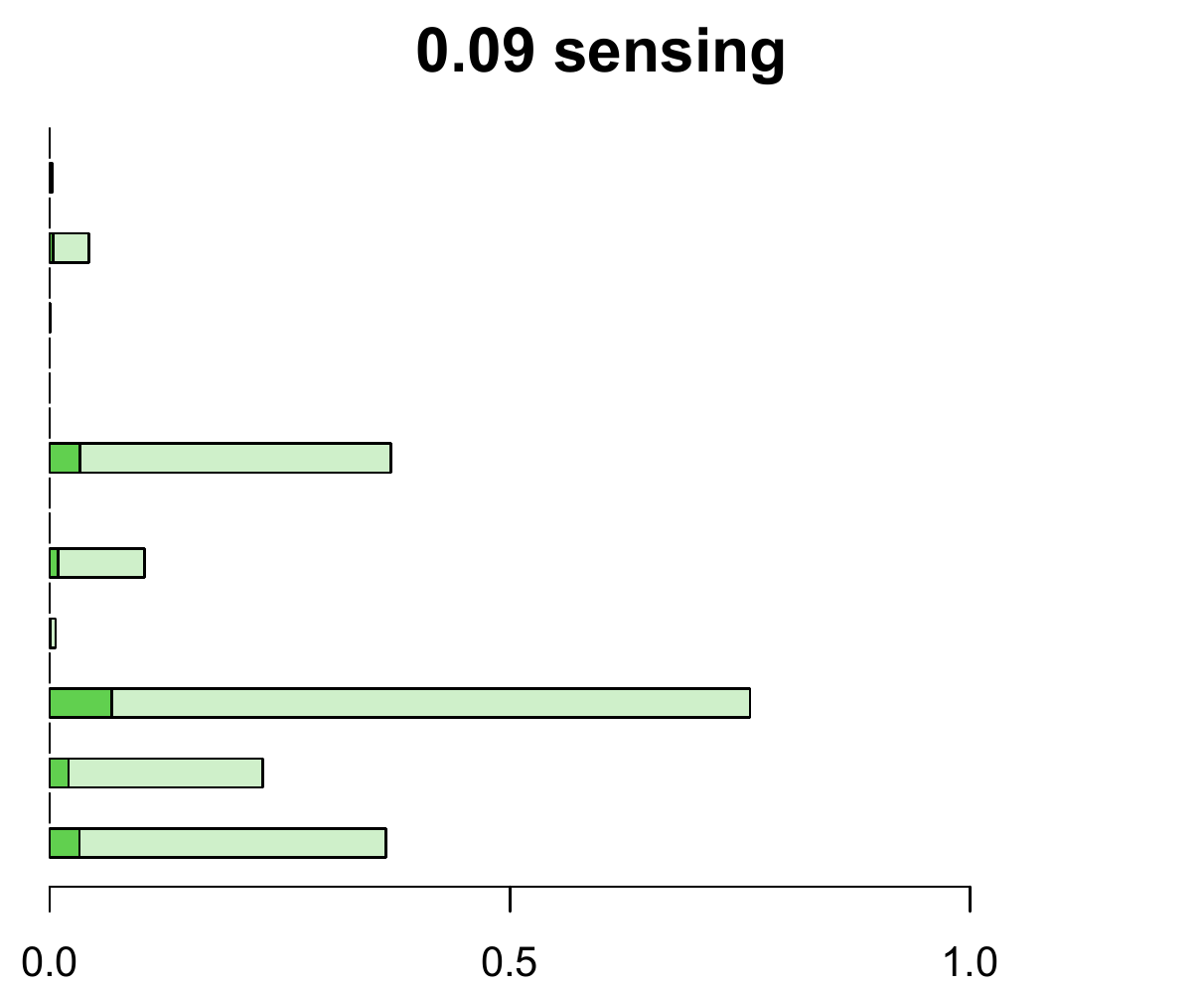}\includegraphics[scale=0.33]{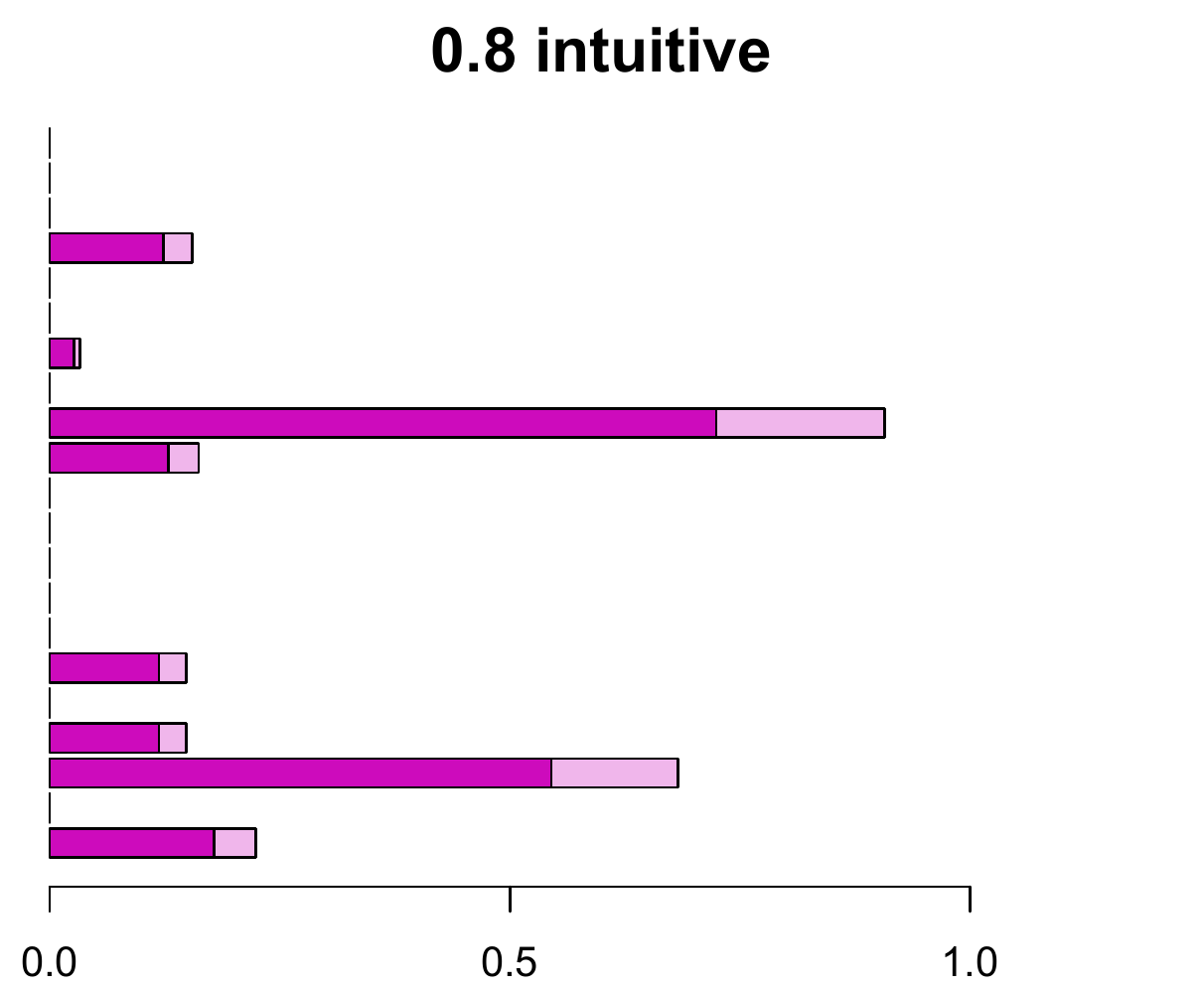}\\
\includegraphics[scale=0.33]{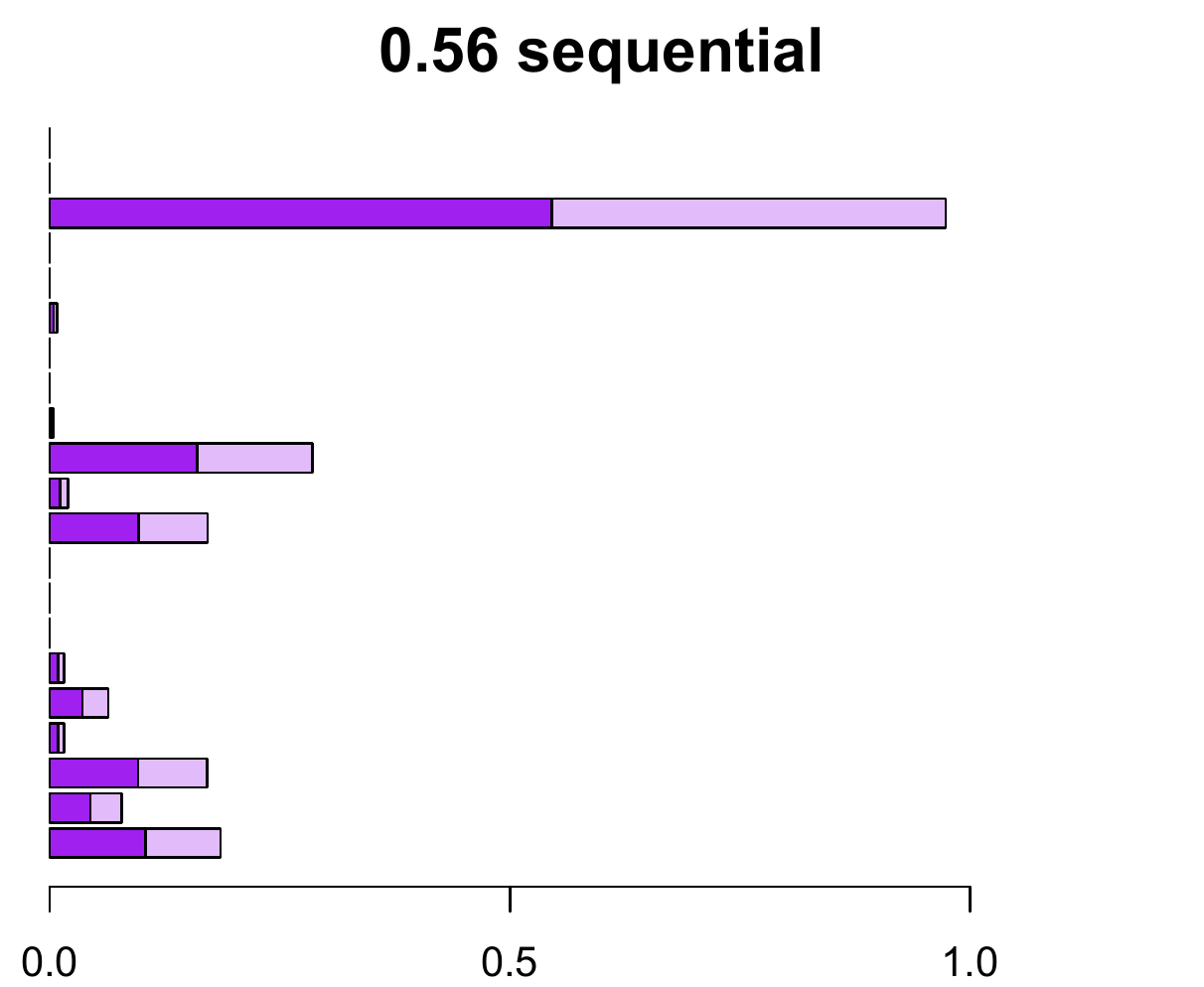}\includegraphics[scale=0.33]{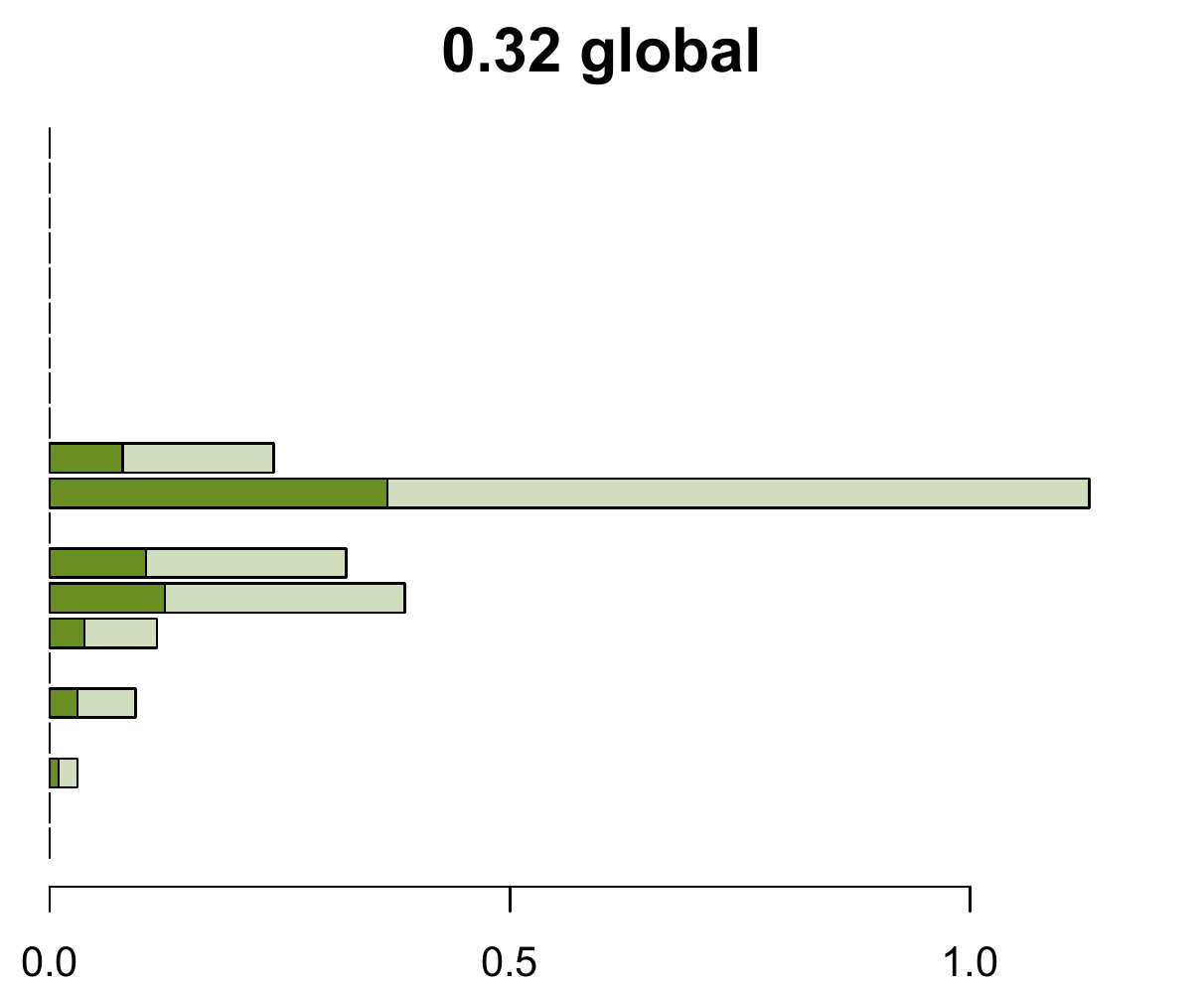}
\includegraphics[scale=0.33]{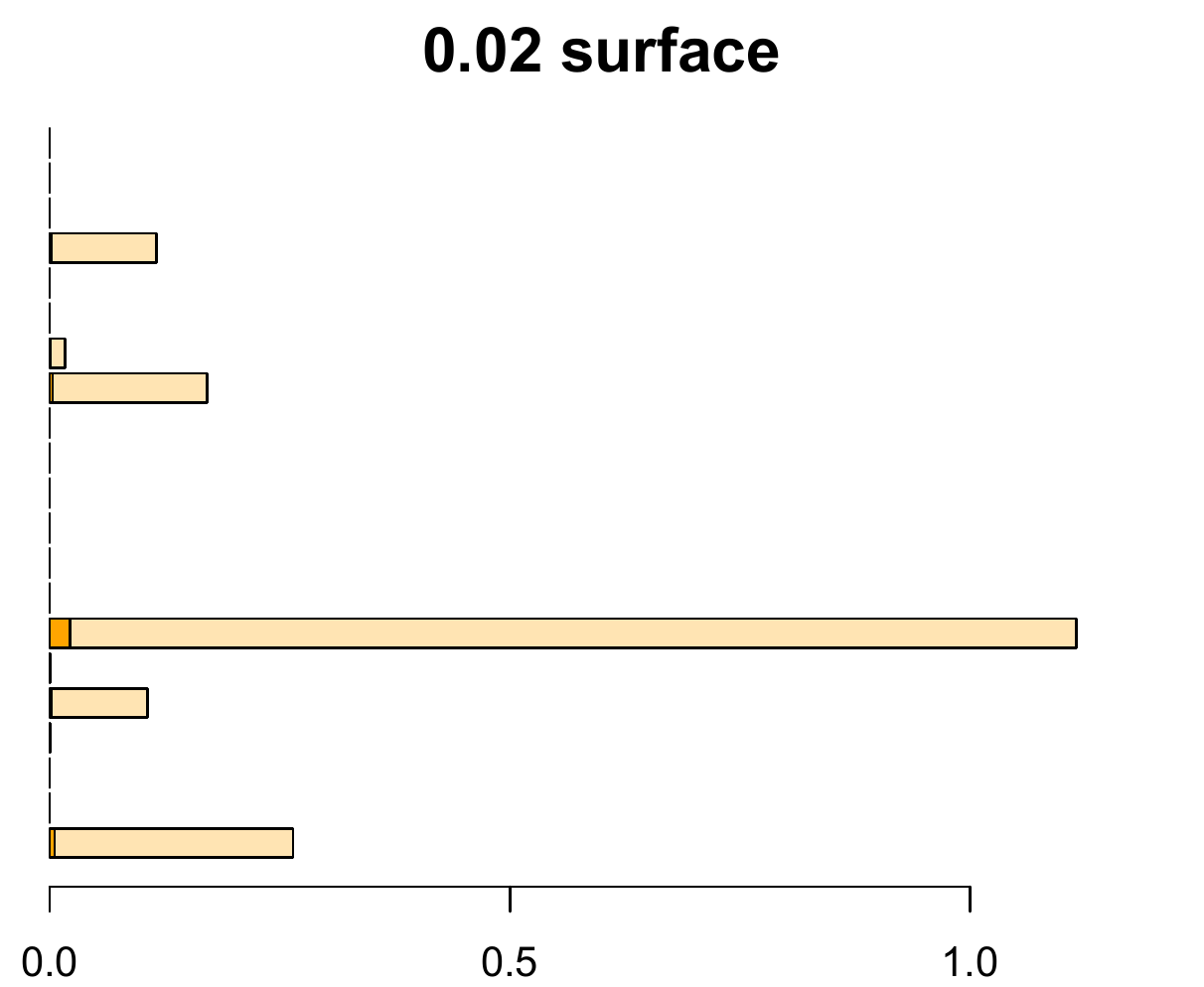}\includegraphics[scale=0.33]{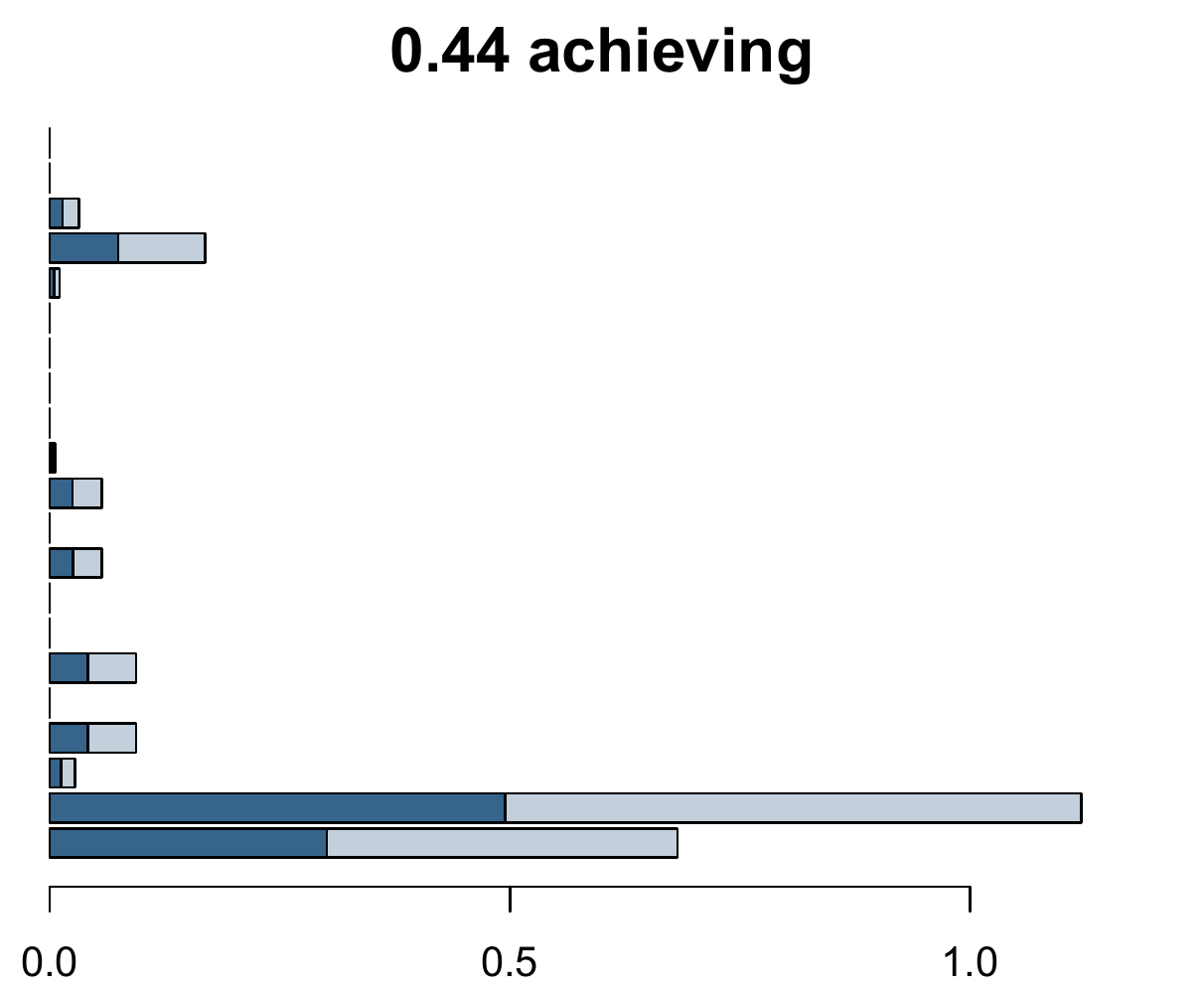}\\

	\caption{\scriptsize{Example 2. Top panel shows the original (left) and the modeled behavior (right) as $X_{\cdot,2}\approx 0.34 P_{\cdot,1} + 0.11 P_{\cdot,2}  + 0.09P_{\cdot,3}  + 0.8P_{\cdot,4}  + 0.56P_{\cdot,5} + 0.32 P_{\cdot,6}  + 0.02  P_{\cdot,7}  + 0.44  P_{\cdot,8}.$  Bottom panels show the learning patterns (light bars) scaled by the affinities (dark bars).
			}}
\label{ex2}
\end{figure}

\vspace{-0.5cm}
Let's also inspect the entries in $A$  representing the individual affinities. Their summary statistics for each pattern are given in Table \ref{affs}. They reveal, that numerous learners have zero affinity to either active, reflective, global, or surface styles. On the other side, most learners show positive affinity to sensing, intuitive, sequential and achieving patterns. Especially, affinities towards the intuitive pattern are far from zero for many learners, which could indicate, that this pattern reflects a combination of a general factor  such as "understanding of theoretical concepts" and the actual intuitive pattern. The later suggests, that to identify an intuitive learner, one have to observe a very high affinity value towards the pattern. 
\begin{table}[ht]
\centering
\small
\begin{tabular}{rrrrrr}
  \hline
pattern & $k$&$\hat q_{k,0.25}$ & $\hat\mu_k$ &$\hat q_{k,0.5}$&$\hat q_{k,0.75}$ \\ 
  \hline
active &     1 & 0.0001 & 0.1044 & 0.0559 & 0.1282 \\ 
  reflective &     2 & 0.0000 & 0.0988 & 0.0105 & 0.0917 \\ 
  sensing &     3 & 0.0876 & 0.2617 & 0.2178 & 0.3852 \\ 
  intuitive &     4 & 0.5921 & 0.6270 & 0.7429 & 0.8250 \\ 
  sequential &     5 & 0.1987 & 0.4552 & 0.4819 & 0.6464 \\ 
  global &     6 & 0.0000 & 0.1704 & 0.0576 & 0.2769 \\ 
  surface &     7 & 0.0036 & 0.1615 & 0.0724 & 0.2148 \\ 
  achieving &     8 & 0.1320 & 0.3233 & 0.2663 & 0.5193 \\ 
   \hline
\end{tabular}
\caption{\scriptsize Summary statistics for the obtained learning pattern affinities in $A$. $\hat q_{k,p}$ denotes the empirical $p$-quantile of the $k$th pattern affinities and $\mu_k$ denotes their mean.}
\label{affs}
\end{table}
\begin{figure}[H]
	\centering
	\includegraphics[scale=0.6]{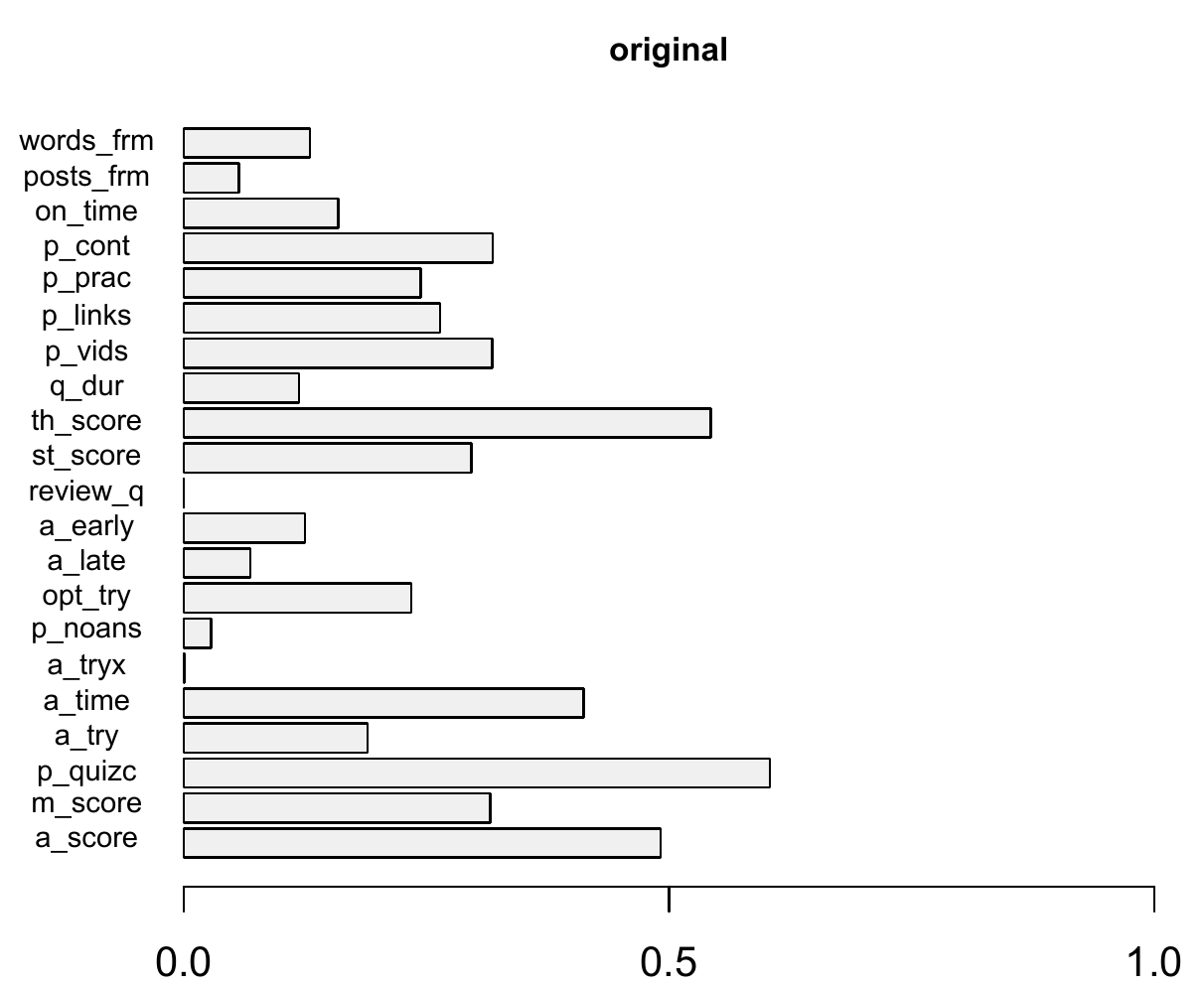} \includegraphics[scale=0.6]{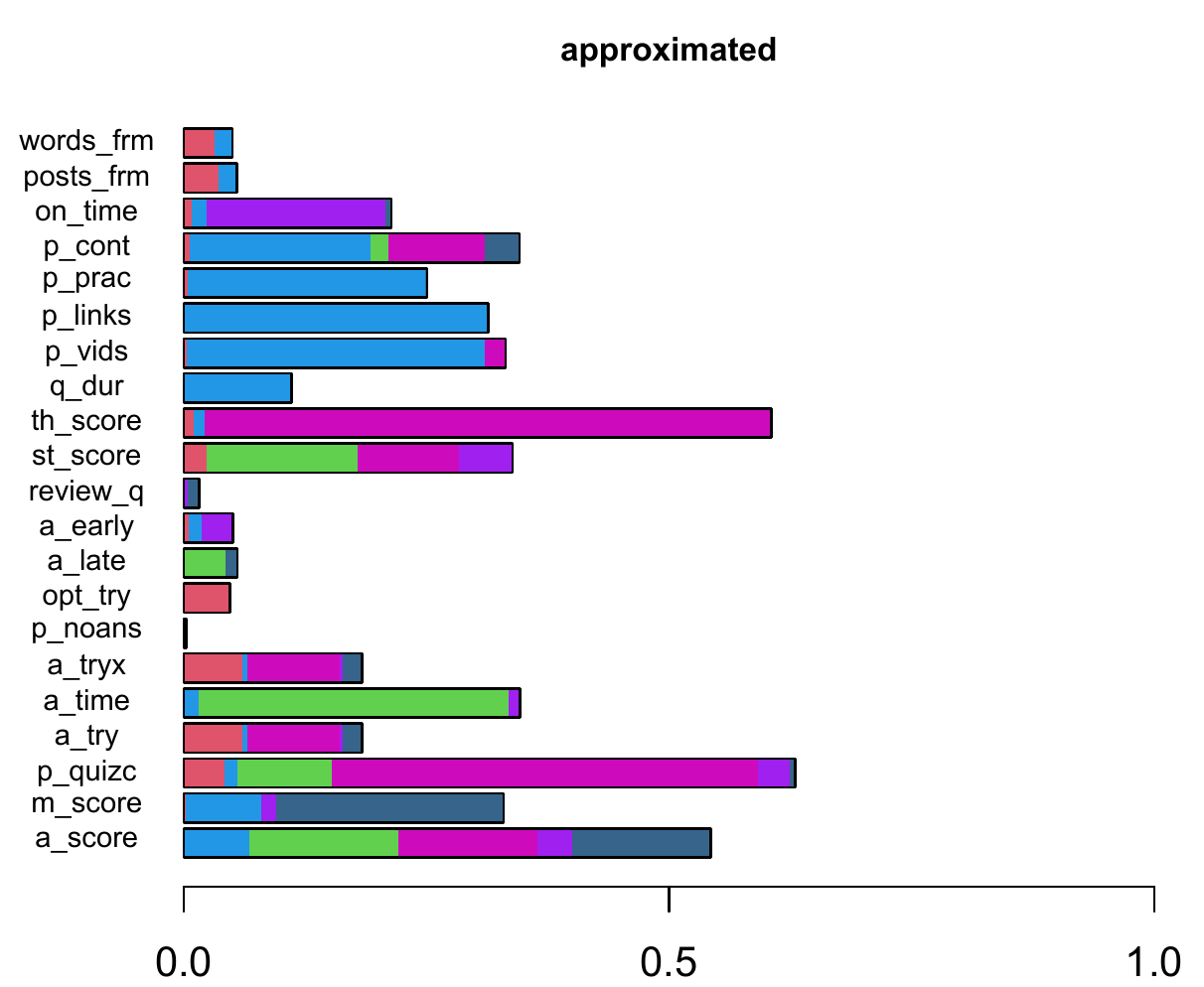} \\
\vspace{0.2cm}
\includegraphics[scale=0.33]{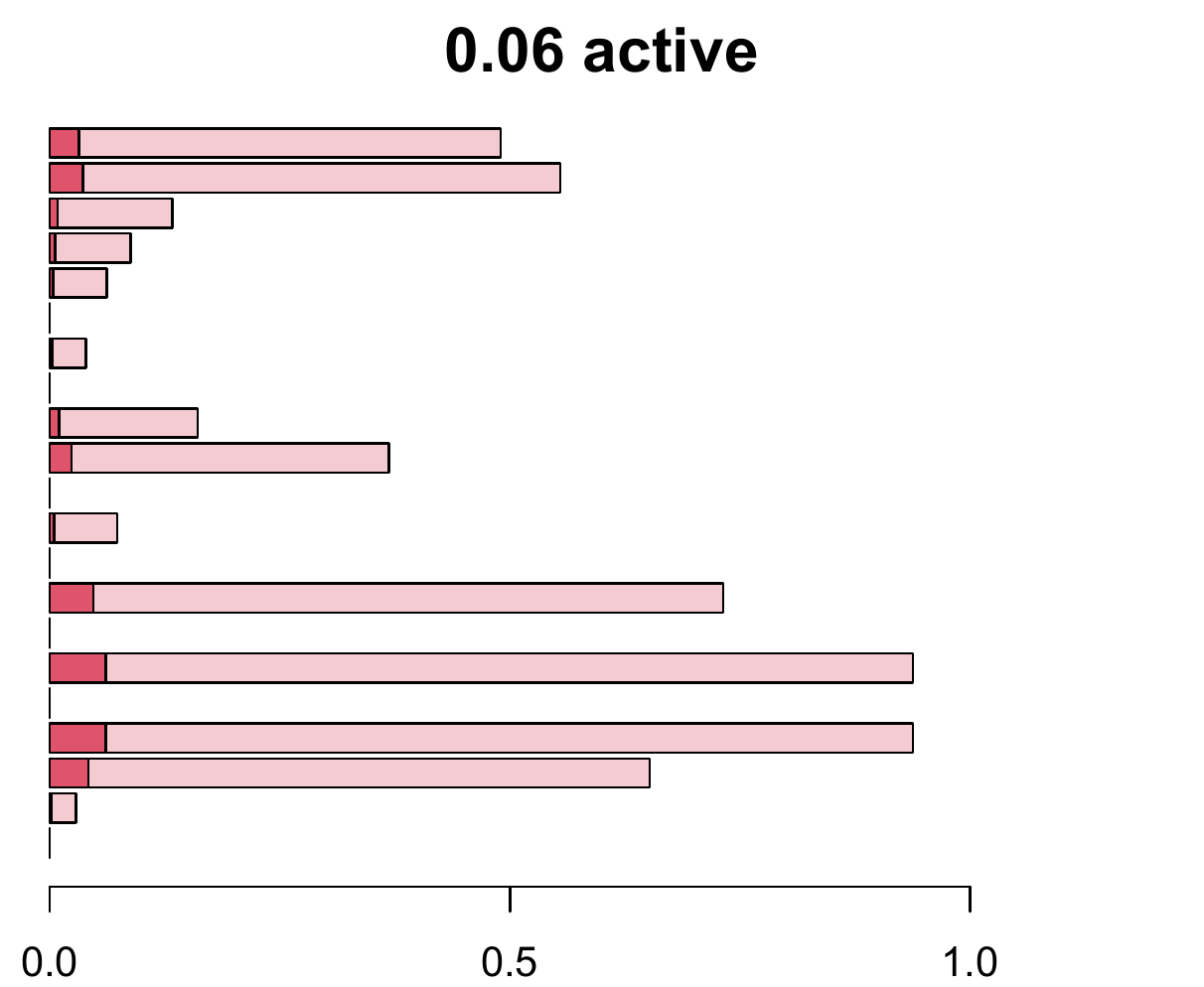}\includegraphics[scale=0.33]{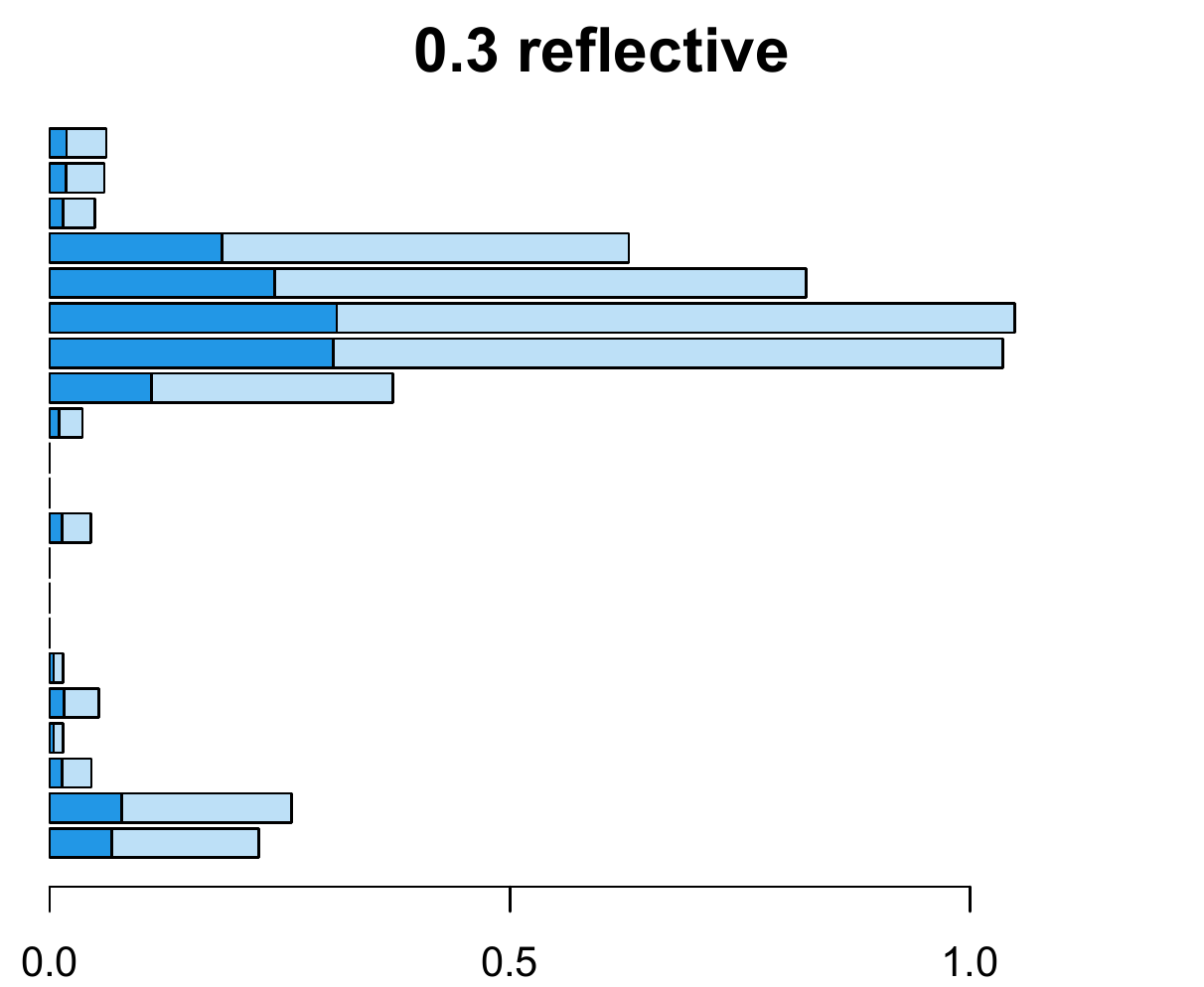}
\includegraphics[scale=0.33]{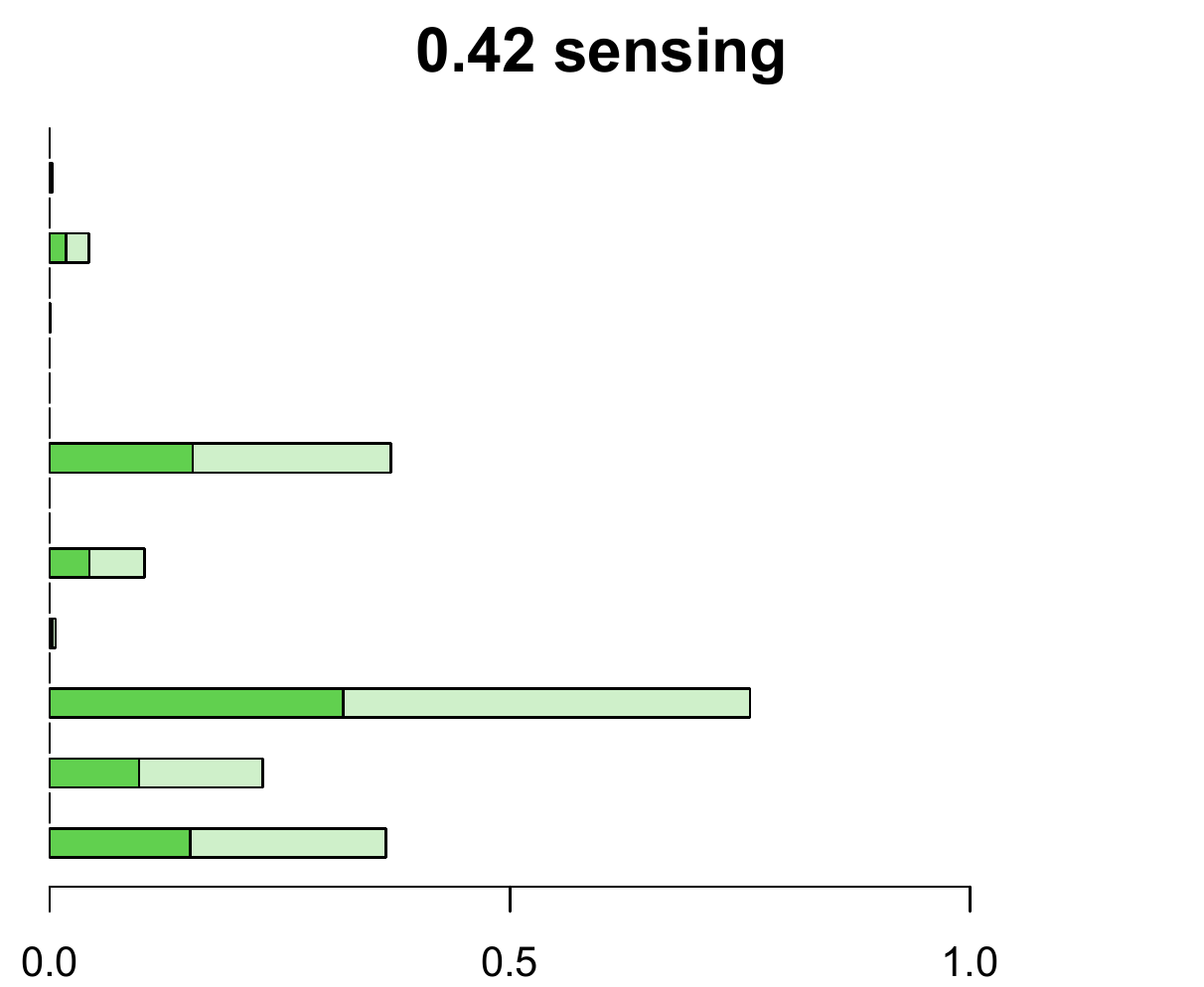}\includegraphics[scale=0.33]{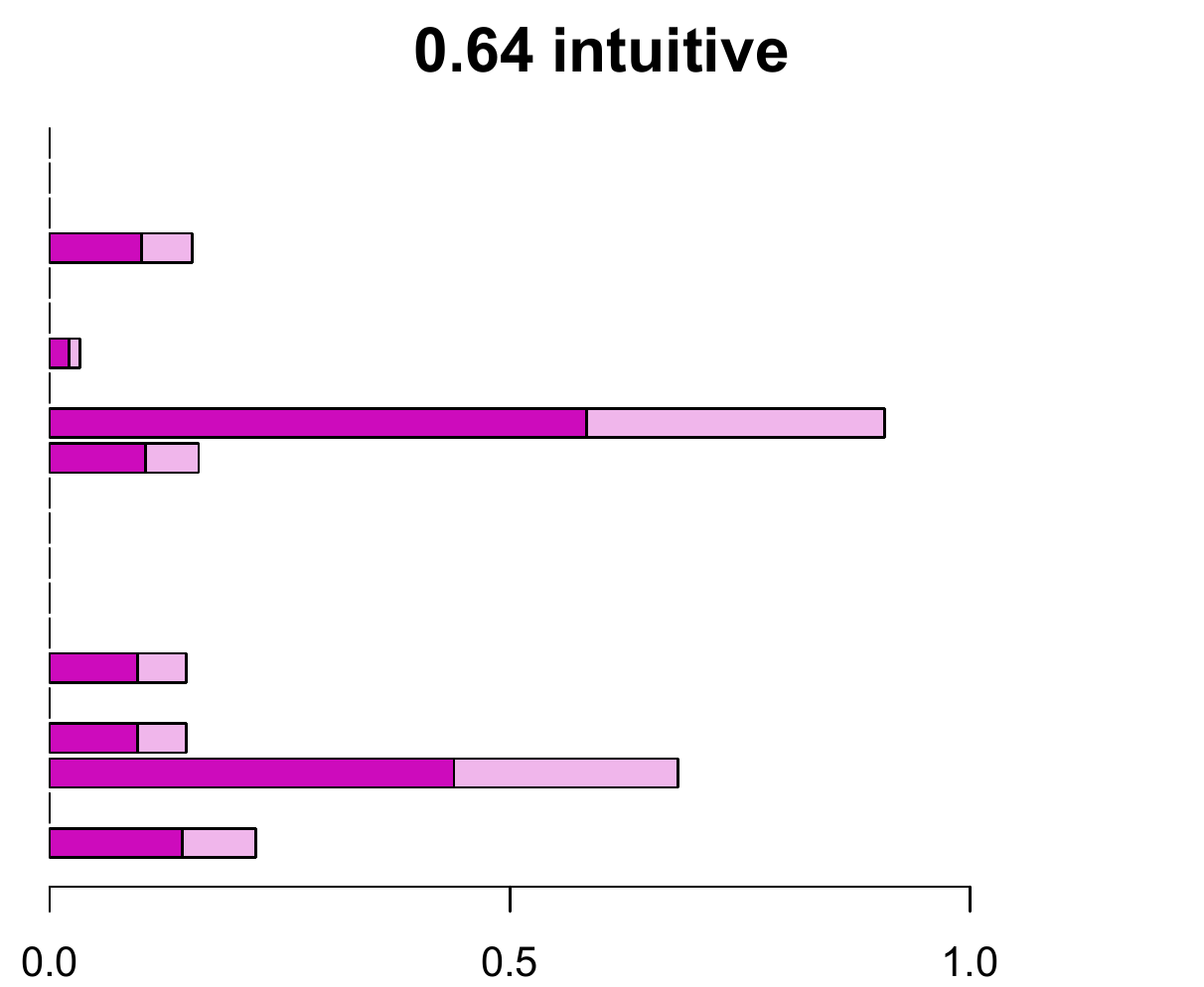}\\
\includegraphics[scale=0.33]{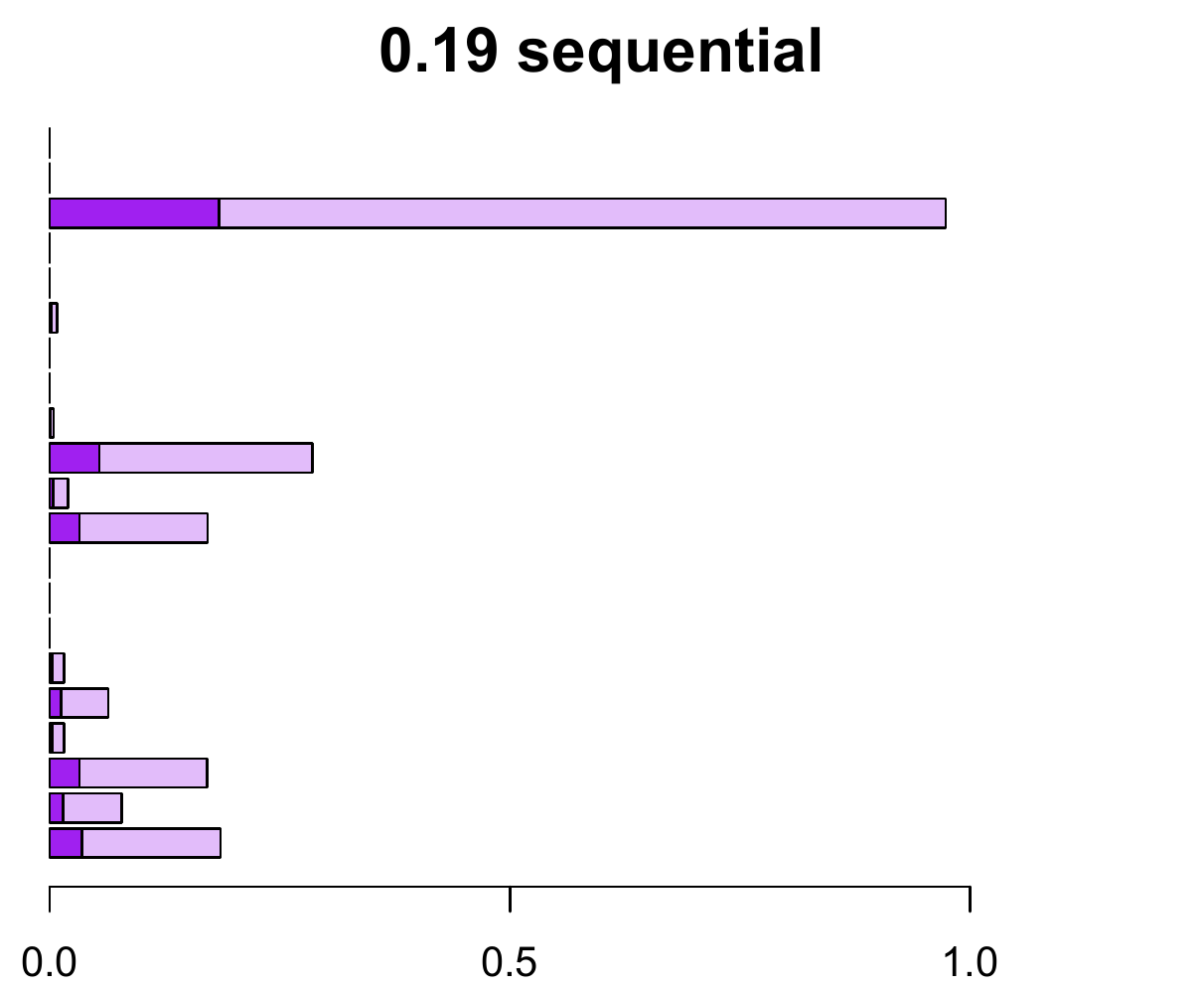}\includegraphics[scale=0.33]{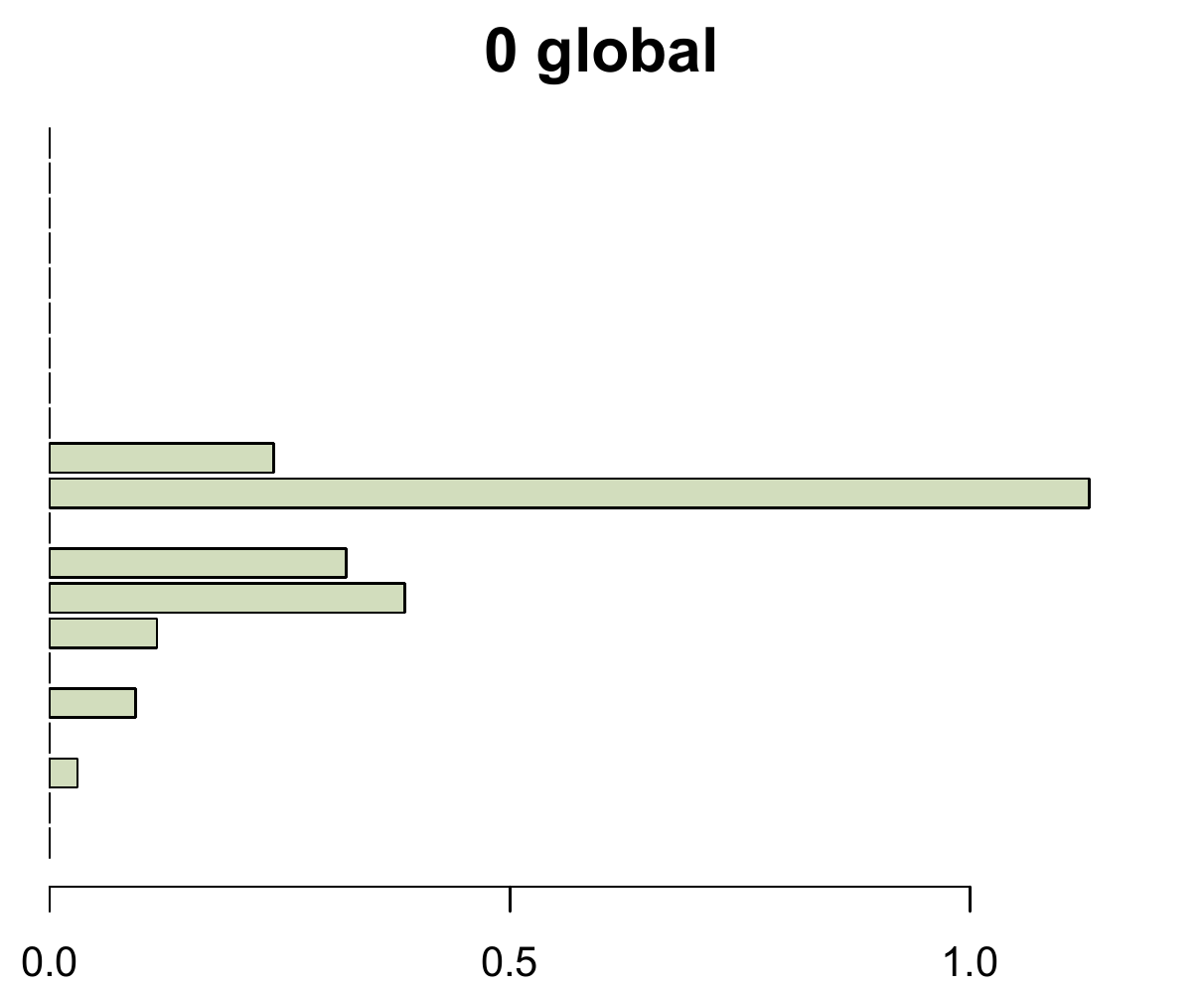}
\includegraphics[scale=0.33]{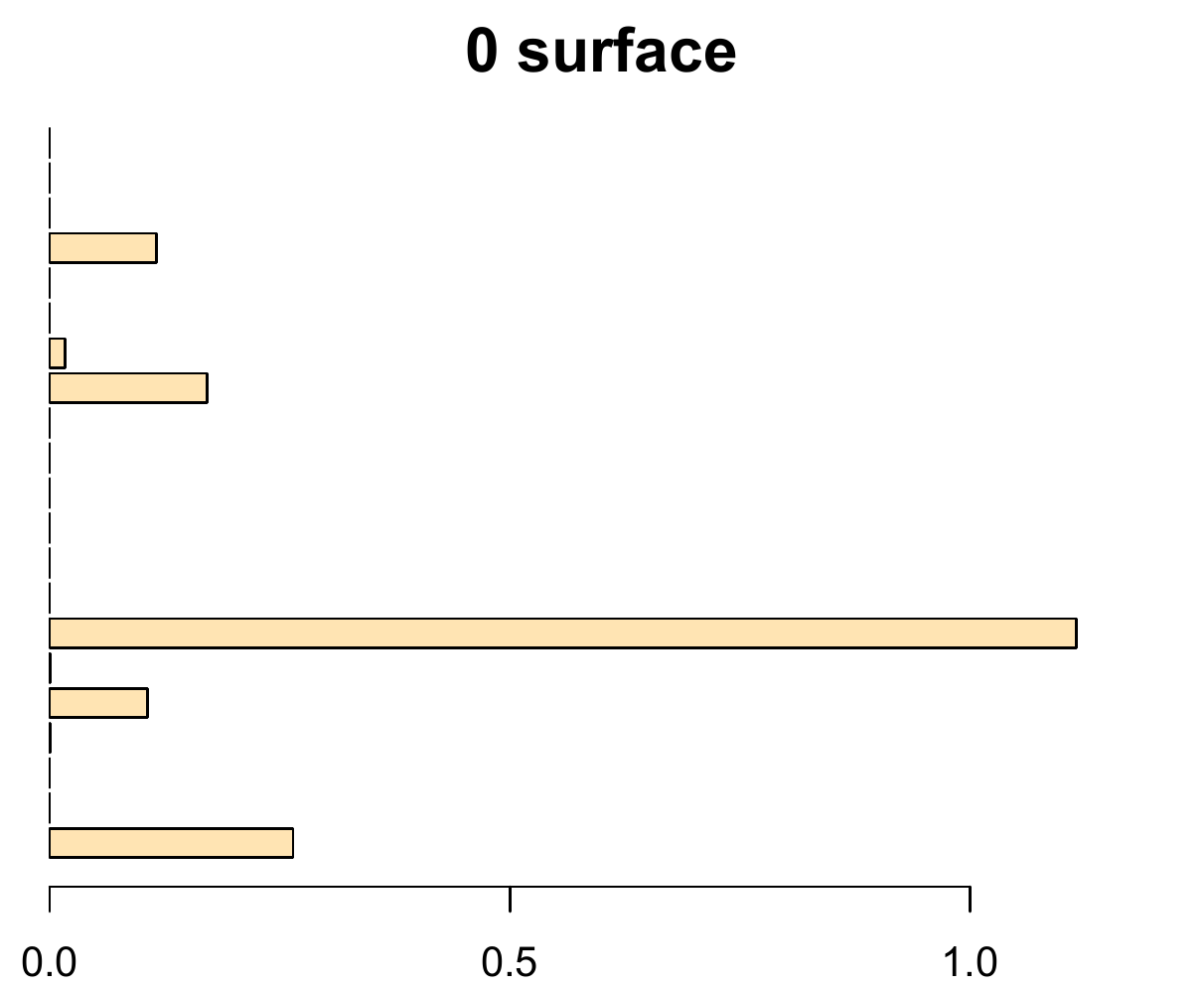}\includegraphics[scale=0.33]{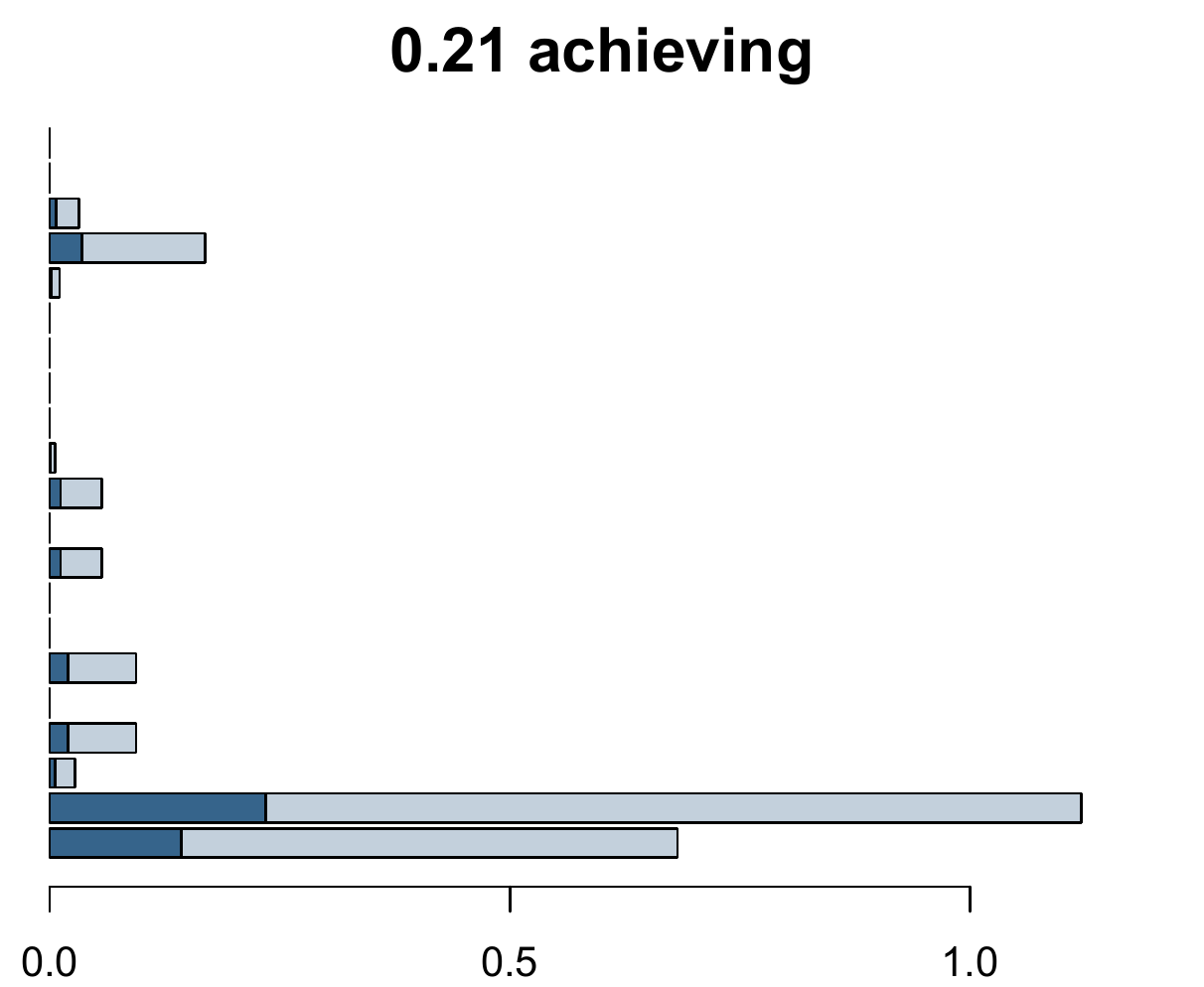}\\
\caption{\scriptsize{Example 1. Top panel shows the original (left) and the modeled behavior (right) as $X_{\cdot,1}\approx 0.06 P_{\cdot,1} + 0.3 P_{\cdot,2}  + 0.42 P_{\cdot,3}  + 0.64P_{\cdot,4}  + 0.19P_{\cdot,5} + 0  P_{\cdot,6}  + 0  P_{\cdot,7}  + 0.21  P_{\cdot,8}.$  Bottom panels show the learning patterns (light bars) scaled by the affinities (dark bars).
			}}
\label{ex1}
\end{figure}
The behavior of each learner is now approximately obtained by adding up the  pattern coefficients multiplied by the respective affinities as written in (\ref{xx}) and visualized in Figures (\ref{ex1}) and (\ref{ex2}).
Moreover, since patterns in $P$ are common to all considered learners, each of them can be sufficiently represented by the affinities.  Instead of considering all 21 features for assessing their learning behavior, we can use the corresponding 8 affinities for exploring the data e.g. for clustering and visualization. 
Consider the following two constructed examples of learning behavior and its approximation through the learning patterns and the affinities in Figures \ref{ex1} and \ref{ex2}.
 In the first example, the learner is represented by the vector of affinities $(0.06,0.3,0.42,0.64,0.19,0,0,0.21)^\top$. In the second example, another vector of affinities \\ $(0.34,0.11,0.09,0.8,0.56,0.32,0.02,0.44)^\top$ is relevant. To compare the two learners, one can now compare their affinities to the learning patterns. For example, the first learner appears to follow the reflective learning strategy most of the course (the affinity of 0.3 is above average for $k=2$, see Table \ref{affs}), the second learner shows more inclination towards the active style. In the first case, the sensing style is more pronounced. In the second case, the affinity for the intuitive style is above average, whereas the sensing one falls below. As a consequence, one would rather count the first learner to the sensing and the second one to intuitive learners.

\subsection{Checking the learning style balance of the course}
A {\it learning style balanced} course describes an instruction, where learners of all learning styles are addressed, which is a leading idea of course improvement  in \cite{felder2} \footnote{ \cite{felder2} speaks of "balancing instruction among different learning style preferences" and stresses that "instruction should routinely address all categories of a selected learning style model" to reach such a balance.}. Under the assumption that learning styles are independent of learners ability\footnote{The assumption of independence between learning styles and ability is quite restrictive for such learning patterns as achieving or surface and may not hold in general. Also, intuitive style seem to combine such factor as "general understanding of theoretical concepts" and actual intuitive learning style, of which the first one is likely to be related to the individual ability.}, one can explore, whether some learners failed to succeed just because of missing suitable learning tools. Therefore, we divide all learners in two groups "p" (passed) and "f"  (failed) based on their performance in the final examination. The groups are somewhat unbalanced: the "f"-group constitutes about 27\% of all students.
\begin{table}[ht]
\centering
\begin{tabular}{rrrrr}
  \hline
pattern &$ k$ & $\hat\mu_{k,f} $&$\hat\mu_{k,p} $& $\hat\sigma_{k,pool} $ \\ 
  \hline
active &     1 & 0.0796 & \bf 0.1136 & 0.1538 \\ 
  reflective &     2 & 0.0648 & \bf 0.1114 & 0.1974 \\ 
  sensing &     3 & \bf 0.3496 & 0.2292 & 0.2135 \\ 
  intuitive &     4 & 0.5663 &\bf 0.6495 & 0.3057 \\ 
  sequential &     5 & 0.4175 & \bf 0.4691 & 0.2734 \\ 
  global &     6 & 0.1651 &\bf 0.1723 & 0.2356 \\ 
  surface &     7 &\bf  0.2009 &  0.1470 & 0.2163 \\ 
  achieving &     8 & 0.1716 &\bf 0.3794 & 0.2298 \\ 
   \hline
\end{tabular}
\caption{\scriptsize  Means ($\hat\mu_{k,g}$) and pooled standard deviations  ($\hat\sigma_{k,pool}$) of the resulting affinities in matrix $A$ computed separately for the two groups ($g\in\{f,p\}$) with "p" passed and "f" failed in the final examination and for $k=1,\ldots, 8$ learning patterns. Larger means are bold.}
\label{sumaffs}
\end{table}

The resulting means and standard deviations of the per-group and per-pattern affinities are shown in Table \ref{sumaffs}. Comparison of means differences under consideration of the pooled standard deviations in Table\ref{sumaffs} suggests that on average being more active, reflective,  intuitive, or achieving could payoff in final exam score.  And the other way around, being on average less sensing and less affine to surface learning could result in better performance on the final exam. Whereas surface and achieving factors are rather related to personal motivation, the group comparison of sensing / intuitive affinities may indicate possible deficiencies of the learning materials in the course. 

The differences in means give an idea of how different the average affinities are between the two groups. Significantly different means of the two groups for some patterns indicate, that the affinities to these patterns may play an important role in determining whether a learner succeeds in the final examination or not. It is to expect, however, that some patterns as achieving or surface, which have to do with the motivation of a learner, do play an important role, but are rather uninteresting for the course improvement. Other patterns relating to active/ reflective, sensing/intuitive or sequential/ global  (ideally) should not exhibit any significant difference in the mean affinities of the groups in a balanced course.

Using again non-parametric bootstrap (\cite{efron93}) with $B=10^4$ replications, I test the significance of the differences in score means by bootstrapping the data, estimating $P_b$ and $A_b$ for each bootstrap replication $b=1,\ldots, B$, and computing group differences for the mean affinities while randomly permuting the groups. 
\begin{table}[ht]
\centering
\begin{tabular}{lllll}
  \hline
pattern&$k$ &hypothesis set $1 $&hypothesis set $2 $& hypothesis set $3 $ \\ 
   \hline
active&1 & 0.4722 & 0.7712 & 0.2288 \\ 
reflective&  2 & 0.3974 & 0.8086 & 0.1914 \\ 
    sensing &3&\bf  0.0426 ** & \bf 0.0228 ** & 0.9772 \\ 
    intuitive&4 &\bf  0.0801 *& 0.9542 &\bf 0.0458 ** \\ 
   sequential& 5 & 0.3964 & 0.7952 & 0.2048 \\ 
  global &  6 & 0.8574 & 0.5976 & 0.4024 \\ 
  surface&  7 & 0.1523 &\bf  0.0660 * & 0.9340 \\ 
  achieving&  8 &\bf 0.0013 ***& 0.9990 &\bf  0.0010 ***\\ 
   \hline
\end{tabular}
\caption{\scriptsize p-values for the three hypotheses sets obtained by bootstrapping the difference in means with $B=10^4$ bootstrap replications;  p-values lower that 10\% are given in bold, statistical significance is indicated by * (10\%), **(5\%), ***(1\%). }
\label{pvals}
\end{table}

Consider the following possible hypothesis sets and each $k=1, \ldots, 8$:
\begin{enumerate}
\item $H_0: \mu_{k,f}=\mu_{k,p}$ against $H_A: \mu_{k,f}\not=\mu_{k,p}$
\item  $H_0: \mu_{k,f}\leq\mu_{k,p}$ against $H_A: \mu_{k,f}>\mu_{k,p}$
\item  $H_0: \mu_{k,f}\geq\mu_{k,p}$ against $H_A: \mu_{k,f}<\mu_{k,p}$
\end{enumerate}

$\mu_{k,g}$  denotes the population mean of the affinities of group $g\in \{f,p\}$ to learning pattern $k=1,\ldots, 8$.
The resulting p-values are reported in Table \ref{pvals}. The differences in means appear statistically significant for the patterns sensing ($k=3$, hypothesis sets 1,2), intuitive ($k=4$, hypothesis sets 1,3), surface ($k=7$, hypothesis sets 2), and achieving ($k=8$, hypothesis sets 1,3). Especially the result for $k=3$ and hypothesis set 2 indicates the need for course improvement concerning the sensing learners.

I conclude, that I should try to re-conceptualize the course to improve the support of sensual learners, e.g. by integrating more real world examples with practical applications, providing some guidance for solving also non-standard problems, and encouraging to concentrate on understanding and not memorizing. 

\section{Conclusions}

In this paper, I propose a data-driven model where individual course interactions are linear combinations of common stylized learning patterns personalized through individual affinities. The model allows to coherently approximate the observed behavior ($X$)  as composed of "building blocks" learning patterns  ($P$) absorbed in the actual data through individual affinities to each pattern ($A$). Using non-negative matrix factorization of $X$, the unknown parts $P$ and $A$ are found preserving the non-negativity of both, which is useful for their interpretation. The setup can be easily adjusted to another learning style system and to any available course interaction data by computing suitable features and changing the number of the components ($K$) depending on which stylized behavior can be captured by the present features. The researcher can explore the resulting learning patterns by relating them to the learning styles and checking their stability through confidence interval construction using bootstrap. The resulting affinities represent learners behavior in lower dimensions and facilitate further usage e.g. for visualizations and clustering. Moreover, using the affinities, a statistical test for differences in means of distinct groups based on their performance (e.g. "failed" and "passed") is available. The results are useful for ensuring a learning style balanced instruction.

\bibliographystyle{apa}
%\bibliography{literature}

\end{document}